\documentclass{article}

\usepackage{PRIMEarxiv}

\usepackage[utf8]{inputenc} 
\usepackage[T1]{fontenc}    
\usepackage{hyperref}       
\usepackage{url}            
\usepackage{booktabs}       
\usepackage{amsfonts}       
\usepackage{amsmath}
\usepackage{nicefrac}       
\usepackage{microtype}      
\usepackage{lipsum}
\usepackage{fancyhdr}       
\usepackage{graphicx}       
\usepackage{natbib}
\usepackage{subcaption} 
\usepackage{microtype}      
\usepackage{xcolor}         
\usepackage{ulem}

\graphicspath{{media/}}     

\renewcommand{\cite}{\citep}
  
\title{
Monkey Perceptogram: Reconstructing Visual Representation and Presumptive Neural Preference from Monkey Multi-electrode Arrays
}

\author{
  Teng Fei\textsuperscript{1}, Srinivas Ravishankar\textsuperscript{1}, Hoko Nakada\textsuperscript{5},  Abhinav Uppal\textsuperscript{4}, Ian Jackson\textsuperscript{1}, \\
  \textbf{Garrison W. Cottrell\textsuperscript{3}, Ryusuke Hayashi\textsuperscript{5}, Virginia R. de Sa\textsuperscript{1,2}} \\
  ${}^{\text{1}}$ Department of Cognitive Science, \\
  ${}^{\text{2}}$ Halıcıoğlu Data Science Institute (HDSI), \\
  ${}^{\text{3}}$ Department of Computer Science and Engineering, \\
  ${}^{\text{4}}$ Shu Chien - Gene Lay Department of Bioengineering, \\
  University of California, San Diego (UCSD) \\
  ${}^{\text{5}}$ Neuroscience Technology Research Group, \\
  Human Informatics and Interaction Research Institute, \\
  National Institute of Advanced Industrial Science and Technology (AIST) \\
  \texttt{\{tfei, srravishankar, auppal, ijackson, desa\}@ucsd.edu} \\
  \texttt{\{h.nakada, r-hayashi\}@aist.go.jp} \\
}

\begin{document}
\maketitle

\begin{abstract}
Understanding how the primate brain transforms complex visual scenes into coherent perceptual experiences remains a central challenge in neuroscience. Here, we present a comprehensive framework for interpreting monkey visual processing by integrating encoding and decoding approaches applied to 
two 
large-scale 
spiking datasets recorded from macaque using THINGS images
(THINGS macaque IT Dataset (TITD) 
and THINGS Ventral Stream Spiking Dataset (TVSD)). 
We leverage multi-electrode array recordings from the ventral visual stream--including V1, V4, and inferotemporal (IT) cortex--to investigate how distributed neural populations encode and represent visual information. Our approach employs linear models to decode spiking activity into multiple latent visual spaces (including CLIP and VDVAE embeddings) and reconstruct images using state-of-the-art generative models. We further utilize encoding models to map visual features back to neural activity, enabling visualization of the "preferred stimuli" that drive specific neural ensembles. Analyses of both datasets reveal that it is possible to reconstruct both low-level (e.g., color, texture) and high-level (e.g., semantic category) features of visual stimuli from population activity, with reconstructions preserving key perceptual attributes as quantified by feature-based similarity metrics. 
The spatiotemporal spike patterns reflect the ventral stream's hierarchical organization with anterior regions representing complex objects and categories. Functional clustering identifies feature-specific neural ensembles, with temporal dynamics show evolving feature selectivity post-stimulus. Our findings demonstrate feasible, generalizable perceptual reconstruction from large-scale monkey neural recordings, linking neural activity to perception.

\end{abstract}


\section{Introduction}

The primate visual system is a model of remarkable computational sophistication, capable of transforming complex, dynamic visual scenes into coherent perceptual experiences and categorical object representations. This transformation is orchestrated by a cascade of cortical areas in the ventral visual stream, from the primary visual cortex (V1) through intermediate regions such as V4, culminating in the inferotemporal cortex (IT), where high-level object and category selectivity emerge \citep{Tanaka1996, Gross1992}. Despite decades of research, the precise neural codes and population dynamics that underlie this transformation remain incompletely understood.

A central challenge in visual neuroscience is to bridge the gap between the high-dimensional, distributed activity of neural populations and the perceptual and semantic content they encode. Two complementary approaches--encoding (modeling how visual stimuli are represented in neural activity) and decoding (reconstructing or inferring visual stimuli from neural responses)--have emerged as powerful tools for probing the representational structure of the visual system. By leveraging these approaches, researchers can move beyond single-neuron analyses to population-level, spatiotemporal characterizations of visual processing, offering new insights into the neural basis of perception \citep{Kay2011, Wu2006}.

Recent advances in large-scale neural recording technologies have enabled the simultaneous measurement of spiking activity from hundreds to thousands of neurons across multiple cortical areas in awake, behaving monkeys \cite{Papale2025, Hayashi2025}. These technological developments, combined with the availability of rich, naturalistic stimulus sets and sophisticated computational models, set the stage for interpreting the dynamic, distributed neural representation of visual experience in the primate brain.

\subsection{Related Work}

Early studies of primate visual processing focused on characterizing the tuning properties of individual neurons in the ventral stream using simple, controlled stimuli. These foundational experiments revealed that V1 neurons are selective for basic features like orientation and spatial frequency \cite{Hubel1968}, V4 neurons for color and intermediate shapes \cite{Zeki1973, Desimone1987}, and IT neurons for complex objects and categories, including faces and animals \cite{Tanaka1991, Gross1972}.

Advances in large-scale neural recording technologies \cite{Papale2025} and the use of naturalistic stimuli \cite{Felsen2005} have enabled population-level, spatiotemporal characterizations of visual processing. 
The development of deep learning-based encoding models \cite{Cadena2019, Guclu2015} and generative decoding approaches \cite{Bashivan2019, Walker2019} further facilitate the interpretation of neural data.
These models have been instrumental in revealing the hierarchical organization of the ventral stream and in quantifying the feature selectivity of neurons at different stages.

Decoding approaches -- which aim to infer or reconstruct the visual stimulus from neural activity—have also played a crucial role in elucidating the information content of neural codes. Early decoding studies demonstrated that it is possible to classify stimulus identity or category from population activity in IT \cite{Hayashi2018}\cite{Hung2005}, and more recent work has extended these methods to reconstruct images or estimate the semantic content of visual scenes from neural data \cite{takagi_high-resolution_2023}. 

The field has recently benefited from the release of large-scale, high-quality datasets that enable population-level analyses of visual processing. The THINGS Ventral Stream Spiking Dataset (TVSD) is a landmark resource, providing spiking activity from V1, V4, and IT in macaque monkeys in response to over 25,000 natural images spanning 1,854 object categories \cite{Papale2025}. The TVSD dataset enables detailed studies of neuronal tuning, noise correlations, and the hierarchical organization of visual representations, and it facilitates direct comparisons with human neuroimaging data due to the shared use of the THINGS image database \cite{Hebart2019, Hebart2023}.
The TVSD paper demonstrated the use of CNN-based encoding models to predict neuronal responses and to generate "most exciting inputs" (MEIs) that visualize the features driving neural activity. These analyses revealed that V1 neurons are tuned to edges and colors, V4 to textures and curvilinear patterns, and IT to complex objects and semantic categories \cite{Papale2025}.

A major recent advance in the field is the "MonkeySee" paper \cite{conwell2024monkey}, which represents the first published attempt to directly reconstruct images from the TVSD dataset. By applying state-of-the-art decoding models to large-scale spiking data, the MonkeySee study demonstrated the feasibility of reconstructing natural images from population activity in the monkey ventral stream. This work highlights the potential of decoding approaches to reveal the perceptual content encoded by neural populations and sets a new benchmark for future studies in neural decoding and perceptogram analysis.

Recently a new dataset has been collected with more emphasis on the IT cortex. The THINGS macaque IT Dataset (TITD) was designed with a similar scope to the TVSD, but was conducted by an independent research group \cite{Hayashi2025}. The TITD focuses on intensive spike recordings from the IT cortex to investigate hierarchical processing within this area, rather than targeting the entire ventral visual stream. Therefore, using both TITD and TVSD provides a valuable basis for validating the proposed analysis and is expected to reveal visual processing at different scales.  

\subsection{Motivation}

Despite significant progress, several key questions remain unanswered. How are complex visual scenes and object categories dynamically represented across the ventral stream? What is the relationship between the distributed patterns of neural activity and the perceptual content experienced by the animal? 

The motivation for this work is to address these questions by leveraging both encoding and decoding approaches to interpret not only the perceptual content, but also the spatiotemporal map of neural activity underlying visual perception in the monkey brain. By applying these methods to large-scale neural recordings, such as those provided by the 
TITD and 
TVSD dataset, and by building on the foundation of earlier encoding and decoding studies, we aim to synthesize high-dimensional neural data into interpretable representations that capture the evolution of perceptual content across space and time. We will bridge the gap between neural activity and perception by reconstructing visual stimuli from population activity and visualizing the features encoded by neural ensembles.

\section{Methods}

\subsection{Datasets}

\begin{table}[h]
\centering
\caption{Comparison of dataset characteristics across subjects}
\begin{tabular}{lccc}
\hline
\textbf{Parameter} & \textbf{TITD (BM160)} & \textbf{TVSD (Monkey N)} & \textbf{TVSD (Monkey F)} \\
\hline
\multicolumn{4}{l}{\textit{Array Configuration}} \\
Total arrays (all areas) & 4 (IT only) & 15 & 16 \\
Arrays in V1 & - & 7 & 8 \\
Arrays in V4 & - & 4 & 3 \\
Arrays in IT & 4 & 4 & 5 \\
Array geometry & 11$\times$12 (-4 corners) & 8$\times$8 & 8$\times$8 \\
Electrodes per array & 128 & 64 & 64 \\
\hline
\multicolumn{4}{l}{\textit{Stimulus Set}} \\
Training images & 8,640 & 22,248 & 22,248 \\
Training repetitions & 4 & 1 & 1 \\
Test images & 100 & 100 & 100 \\
Test repetitions & 48 & 30 & 30 \\
\hline
\multicolumn{4}{l}{\textit{Presentation Paradigm}} \\
Stimulus duration (ms) & 300 & 200 & 200 \\
Inter-stimulus interval (ms) & 300 & 200 & 200 \\
\hline
\end{tabular}
\label{tab:dataset_comparison}
\end{table}


\subsubsection{TITD}

The dataset consists of 
spiking neural data in response to
8,640 unique training images from the 
THINGS fMRI1 images
with 4 repetitions each, plus 100 test images presented with 48 repetitions\cite{Hayashi2025}. Neural responses were recorded using 4 chronically implanted Utah arrays positioned along the IT cortex from anterior to posterior. Each array had an 11×12 electrode configuration with 
400 $\mu$m inter-electrode spacing and
corner electrodes absent, yielding 128 functional channels per array and 512 recording electrodes total. 
The dataset includes recordings from an adult male macaque (Macaca fuscata, designated as BM160). The electrode array placement was illustrated in Electrode Array Placement Figure of \cite{Hayashi2025}.
Images were presented using a rapid serial visual presentation (RSVP) paradigm, with each image displayed for 300 ms followed by a 300 ms blank interval.

\subsubsection{TVSD}
As described in the original paper by \citet{Papale2025}, the dataset includes recordings from two adult male macaques (Macaca mulatta, designated as Monkey N and Monkey F) using chronically implanted Utah arrays (Blackrock Microsystems). Each array contained 64 electrodes arranged in an 8×8 configuration with 
400 $\mu$m inter-electrode spacing. Monkey N had 15 functional arrays (7 in V1, 4 in V4, 4 in IT), while Monkey F had 16 arrays (8 in V1, 3 in V4, 5 in IT). 
The electrode array placement for each macaque cortex is illustrated in Figure 1A of \cite{Papale2025}.
Neural responses were recorded while monkeys viewed 22,248 unique natural images from the THINGS database for the training set, comprising 1,854 object categories with 12 exemplars and 1 presentation each, plus 100 test images that were repeated 30 times. 

Images were presented for 200 ms followed by a 200 ms inter-stimulus interval, in sequences of four images per trial while monkeys maintained central fixation. 

\subsection{Model Architecture}
There are three primary components in our pipeline: A linear \textit{decoder} from brain space to latent space, a linear \textit{encoder} mapping this decoded latent back into brain space, and a \textit{reconstructer} that generates an image from the decoder's output latent. The input to the pipeline is brain data, and two outputs are produced. These correspond to the reconstructed image and the latent-filtered brain patterns for that image.

\begin{figure*}[t]
    \centering
    \begin{subfigure}[t!]{0.65\textwidth}
        \centering
        \includegraphics[width=\linewidth]{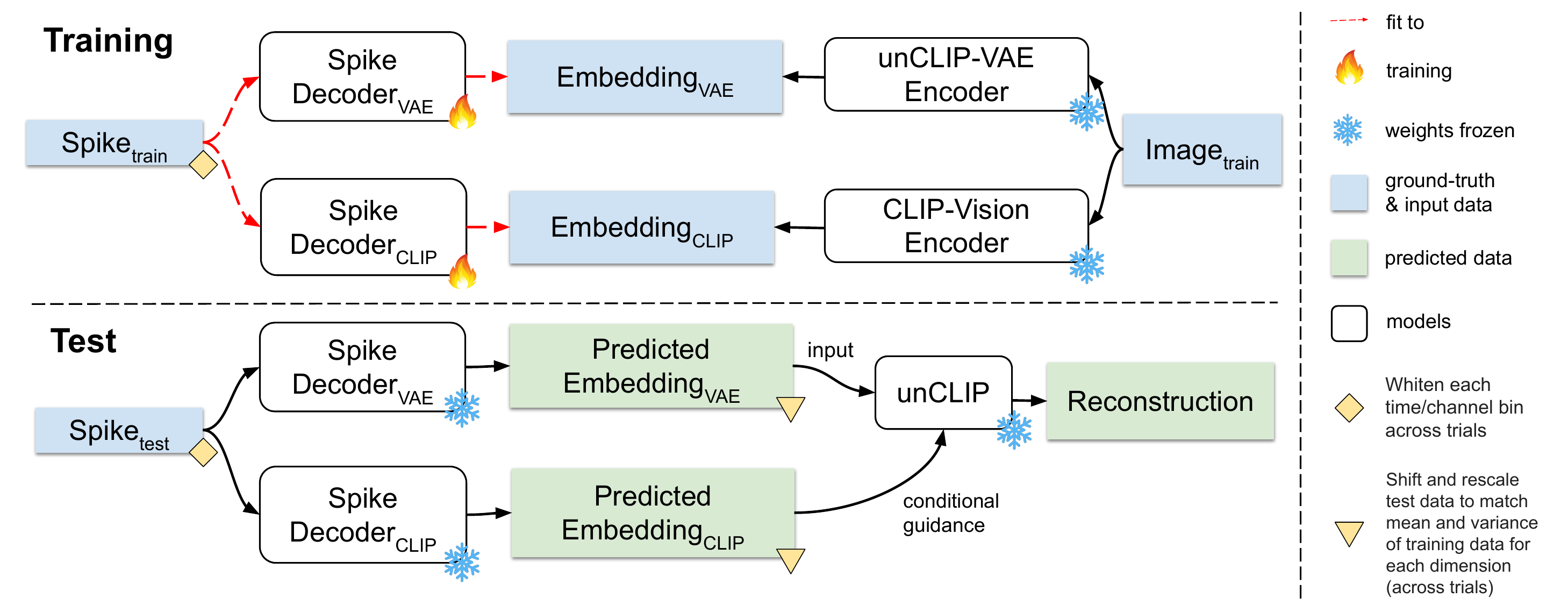}
        \caption{Image-reconstruction pipeline. During training, ridge regressors learn to map spikes into VAE and CLIP latent spaces. Test-time spikes passes through them to predict the latents, which \textit{unCLIP} converts into images.}
        \label{fig:flowchart_unCLIP}
    \end{subfigure}
    \hfill
    \begin{subfigure}[t!]{0.33\textwidth}
        \centering
        \includegraphics[width=\linewidth]{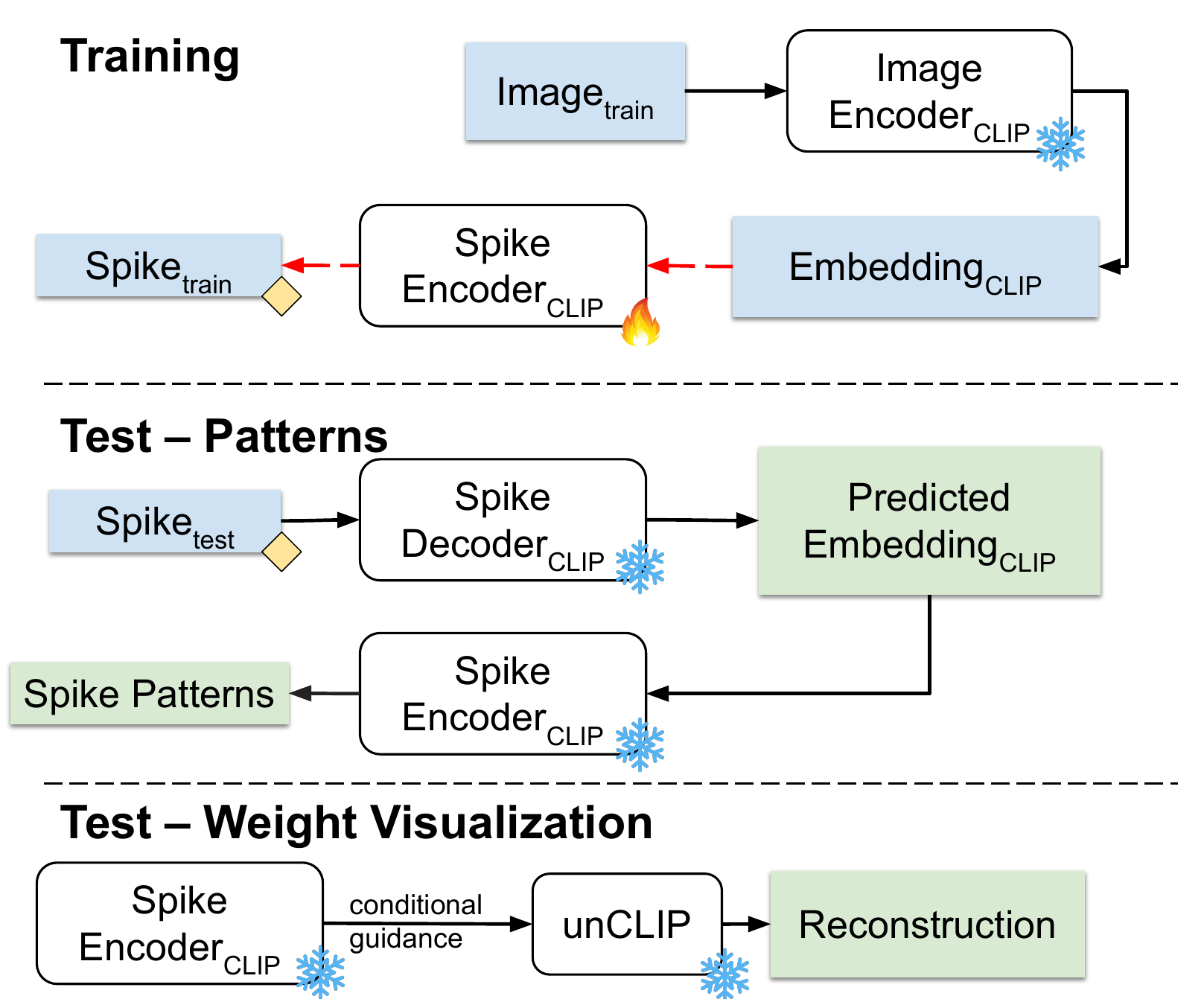}
        \caption{Latent-filtered spike pattern generation and weight visualization. Similar to the left pipeline but trains a CLIP\,$\rightarrow$\,spike encoder to map embeddings back to spikes. Weight visualization treats each channel/time bin of the encoding weight as a CLIP embedding vector and reconstructs it using unCLIP.}
        \label{fig:clip_pattern_flowchart}
    \end{subfigure}

    \caption{Flowcharts for (\textbf{left}) spikes-to-image reconstruction with \textit{unCLIP} and (\textbf{right}) generation of CLIP-filtered spike patterns and encoding weight visualization.}
    \label{fig:unclip_pipeline}
\end{figure*}


\subsubsection{Encoder, Decoder}
We employ linear regression in both the encoder and decoder, shown to be an effective way to decode latents from fMRI \citep{Ozcelik_VanRullen_2023} as well as EEG \cite{Fei2024}. 

\subsubsection{Latent Space}
While Fig. \ref{fig:unclip_pipeline} illustrates the pipeline using the CLIP latent space, we also use other latent spaces which emphasize different visual features in their reconstructions. With CLIP latents, the reconstructions preserve high-level semantic categories of the images. Latents from VDVAE, PCA or ICA might emphasize other lower-level visual features. While ICA reconstructions emphasize color saturation and contrast, PCA reconstructions capture overall brightness well. 

\subsubsection{Reconstructer} 
Reconstructions from CLIP latents require a diffusion module. We have used unCLIP \citep{ramesh_hierarchical_2022}. Images are reconstructed from VDVAE latents using its pretrained frozen decoder, and PCA/ICA simply use linear inverse projections. 

\subsection{Experiments}

\subsubsection{Image reconstruction}
We developed a simple pipeline using unCLIP \citep{ramesh_hierarchical_2022} which we use for most of our qualitative analyses as it provides a simple architecture while maintaining comparable reconstruction performance. A flowchart illustrating image reconstruction for the unCLIP variant of our pipeline is shown in Fig. \ref{fig:unclip_pipeline}.  We introduce an initial VAE encoder into the standard unCLIP framework so our unCLIP diffusion process uses two types of latents as input, an initial (VAE) latent and a conditioning/guidance vector (CLIP).

In the training stage, an image is processed through unCLIP-VAE and CLIP-Vision encoders, producing two target latent embeddings needed for our unCLIP diffuser. Linear decoders are trained using linear regression to map binned spike counts to these targets. Before fitting, the spike data is normalized for each of the channel/time bin dimensions across the 16540 training classes. In the test stage, the decoder predicts corresponding embeddings given a trial's spike data. The predicted embeddings are shifted and rescaled using the mean and variance of the training latent distribution. These are input to the unCLIP decoder to produce a reconstructed image. All pre-trained models are frozen, and the linear decoders are the only modules trained.

\subsubsection{Spike Patterns}


This section describes how we obtain the spatio-temporal spike patterns specific to a visual feature, with textured vs smooth patterns as an example. 
First, we choose a latent space emphasizing the feature of interest, by manually inspecting reconstructions from various latent spaces. Textures appear to be emphasized in reconstructions from VDVAE, as seen in Fig \ref{fig:vdvae_25msbin}.
The encoder and decoder are then trained as described previously. During testing, the decoder is used to first predict image latents from the held-out spike data, and the encoder is used to project these latents back to the spike space, thus `filtering' the spikes through the chosen latent space. This is illustrated in Fig. \ref{fig:clip_pattern_flowchart}, which highlights the decoding-encoding loop during testing. Next, we order the reconstructed images along the visual feature of interest (eg. textured vs smooth images). This is done using heuristics described in the Appendix. Spike patterns for all images in each group are averaged, to form the corresponding group spike pattern (eg. texture pattern vs smooth pattern). Using this procedure, we produce maps for the following visual features: visual semantics (animal vs food vs other), texture (textured vs smooth), hue (red vs blue). 

\subsubsection{Presumptive Neural Preference Visualization via Encoding Weight Reconstruction}

To investigate the semantic tuning properties in our neural-to-CLIP mapping, we performed a detailed analysis of the encoding weights that transform CLIP latent representations into predicted spike counts. These weights provide insight into what visual features maximally drive neural responses at each electrode and time point.


We extracted individual weight vectors from the CLIP-to-spike encoding matrix, where each vector corresponds to a specific electrode at a given time bin. Each weight vector has the same dimensionality as CLIP embeddings (1024 dimensions) and represents the linear transformation from CLIP space to predicted spike counts. To visualize the preferred stimulus features encoded by these weights, we treated each weight vector as if it were a CLIP embedding of an actual image and reconstruct it using unCLIP. This approach reveals the "ideal stimulus" that would maximally activate each electrode according to the learned linear model.

For the TITD dataset (BM160), reconstructed images were organized into spatiotemporal maps spanning 0-600 ms post-stimulus onset in 25 ms bins, corresponding to 24 time points. For effective visualization we only take the first 13 bins corresponding to 0-325 ms. The four horizontal blocks correspond to the four electrode arrays positioned from anterior (top) to posterior (bottom) IT cortex, with electrodes within each block maintaining their physical array positions (11×12 configuration minus corner electrodes, yielding 128 channels per array). 

For the TVSD dataset, we analyzed 10 time points spanning 0-400 ms post-stimulus onset in 40 ms bins. Arrays were displayed in their original recording order: Monkey N with 15 functional arrays (7 in V1, 4 in V4, 4 in IT) and Monkey F with 16 arrays (8 in V1, 3 in V4, 5 in IT), each in an 8×8 configuration. This visualization enables direct observation of how feature complexity and semantic content evolve both temporally (from simple features to complex objects) and spatially (across the anterior-posterior axis of IT).

\subsubsubsection{\textbf{Weight Similarity Analysis and Functional Clustering}}

To identify functional ensembles with similar tuning properties, we analyzed the similarity structure of the encoding weights across all electrodes and time points. Weight vectors from all spatiotemporal locations were aggregated and subjected to hierarchical clustering using cosine similarity as the distance metric. This clustering produced a one-dimensional ordering that groups functionally similar neural sites regardless of their physical location or response timing.

The resulting dendrogram ordering was mapped to a continuous color scale, with each electrode-timepoint combination colored according to its position in the similarity hierarchy. The emergence of contiguous patches of uniform color in the spatiotemporal maps indicates functional ensembles. 



\subsection{Evaluation}
\label{evaluation_metrics}

We used the same performance metrics 
(see Table \ref{tab:performance}) as in \cite{Ozcelik_VanRullen_2023}, which has been used in other followup studies such as MindEye \cite{Scotti_Banerjee_Goode_Shabalin_Nguyen_Cohen_Dempster_Verlinde_Yundler_Weisberg_et_al_2023} and Perceptogram \cite{Fei2024}. The 8 metrics we used are Pixel Correlation (PixCorr), Structural Similarity (SSIM), AlexNet layer 2 and 5 outputs pairwise correlations, InceptionNet output pairwise correlation, CLIP ViT output pairwise correlation, EfficientNet output distance, and SwAV output distance.
PixCorr and SSIM involve comparing the reconstructed image with the ground-truth (GT) test image. PixCorr is a low-level (pixel) measure that involves vectorizing the reconstructed and GT images and computing the correlation coefficient between the resulting vectors. SSIM is a measure developed by Wang et al. 2004 that computes a match value between GT and reconstructed images as a function of overall luminance match, overall contrast match, and a ``structural'' match which is defined by normalizing each image by its mean and standard deviation. 

\section{Results}

\subsection{Reconstruction performance}
The pipeline produces reconstructions consistent with the stimulus images in various aspects such as color, texture, and semantic meaning 
(see Fig. \ref{fig:monkey_n_unclip}). 
The quantitative performance scores of our simple linear model are shown in Table \ref{tab:performance}. Both the quantitative performance and visual quality of the reconstructions are comparable across the two datasets. Features that tend to get well captured tend to be animacy (whether the image is about an animal), color cast (when the image is strongly red, blue, or green), spherical shape, and high texture. 
For our unCLIP reconstructions, having electrodes present in low and mid-level visual areas for Monkey F and Monkey N did not necessarily translate to better performance in capturing lower level visual features such as color and shape compared to Monkey BM160 which only has arrays in IT. However, for our VDVAE reconstructions, Monkey F and Monkey N do capture more visual details compared to Monkey BM160.


\begin{figure}[h]
    \centering
    \includegraphics[width=1\linewidth]{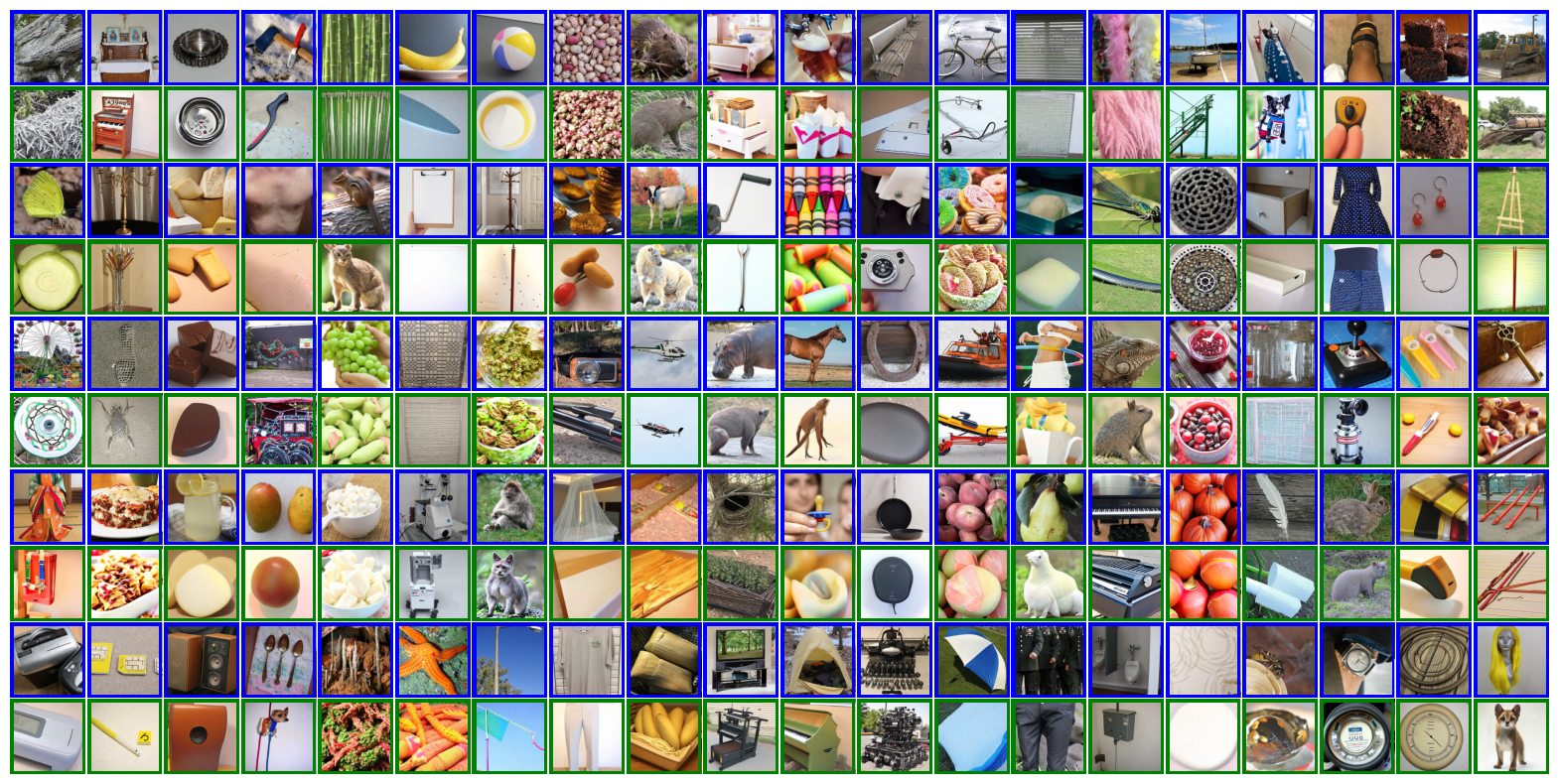}
    \caption{\textbf{Visual reconstruction from TITD Monkey BM160 neural spike data using unCLIP.} Top row in each pair: original stimuli presented to macaque subjects (blue borders). Bottom row: images reconstructed via unCLIP diffusion model (green borders) from CLIP embeddings decoded from multi-unit spike recordings in IT cortex.}
    \label{fig:clip_reconstructions_bm160}
\end{figure}

\begin{figure}[h]
    \centering
    \includegraphics[width=1\linewidth]{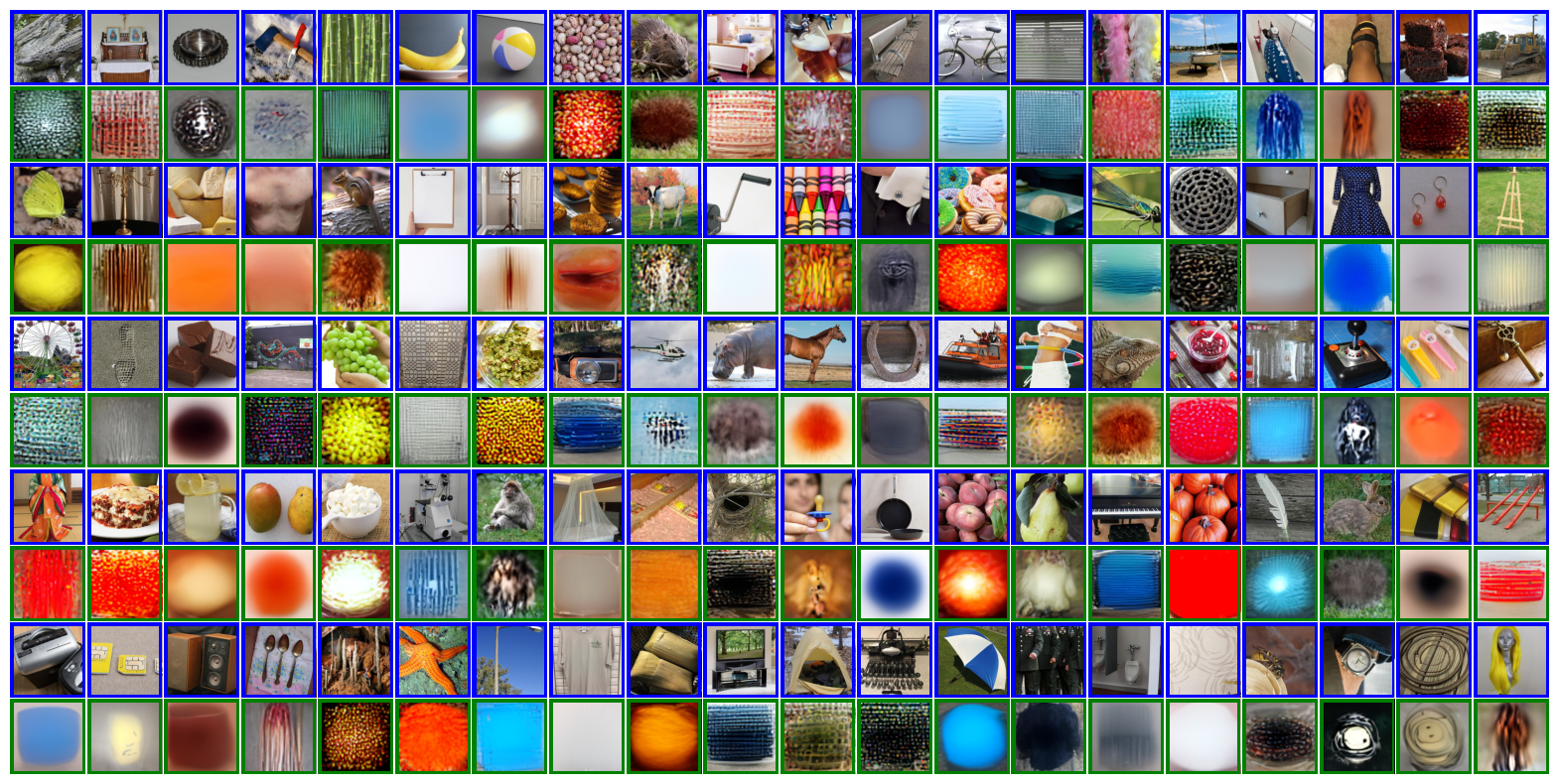}
    \caption{Comparison of ground truth images (blue) with VDVAE reconstructions (green) generated from VDVAE latents decoded from TITD Monkey BM160 IT cortex spike data.}
    \label{fig:vdvae_25msbin}
\end{figure}


\begin{figure}
    \centering
    \includegraphics[width=1\linewidth]{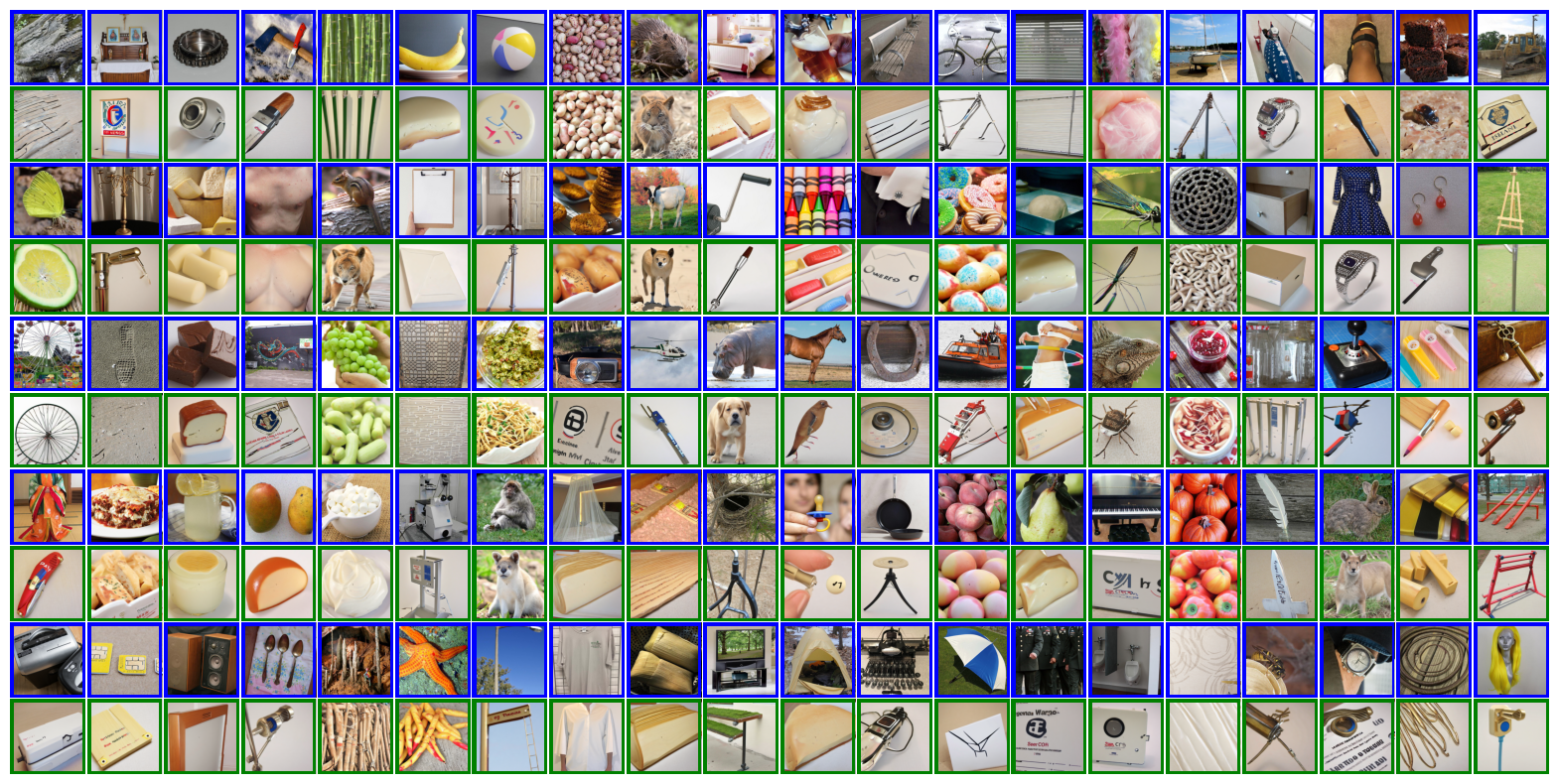}
    \caption{\textbf{Visual reconstruction from TVSD Monkey F neural spike data using unCLIP.} Ground truth images (blue borders) and their corresponding reconstructions (green borders) generated using unCLIP from CLIP embeddings decoded from macaque visual cortex spike recordings.}
    \label{fig:monkey_f_unclip}
\end{figure}

\begin{figure}
    \centering
    \includegraphics[width=1\linewidth]{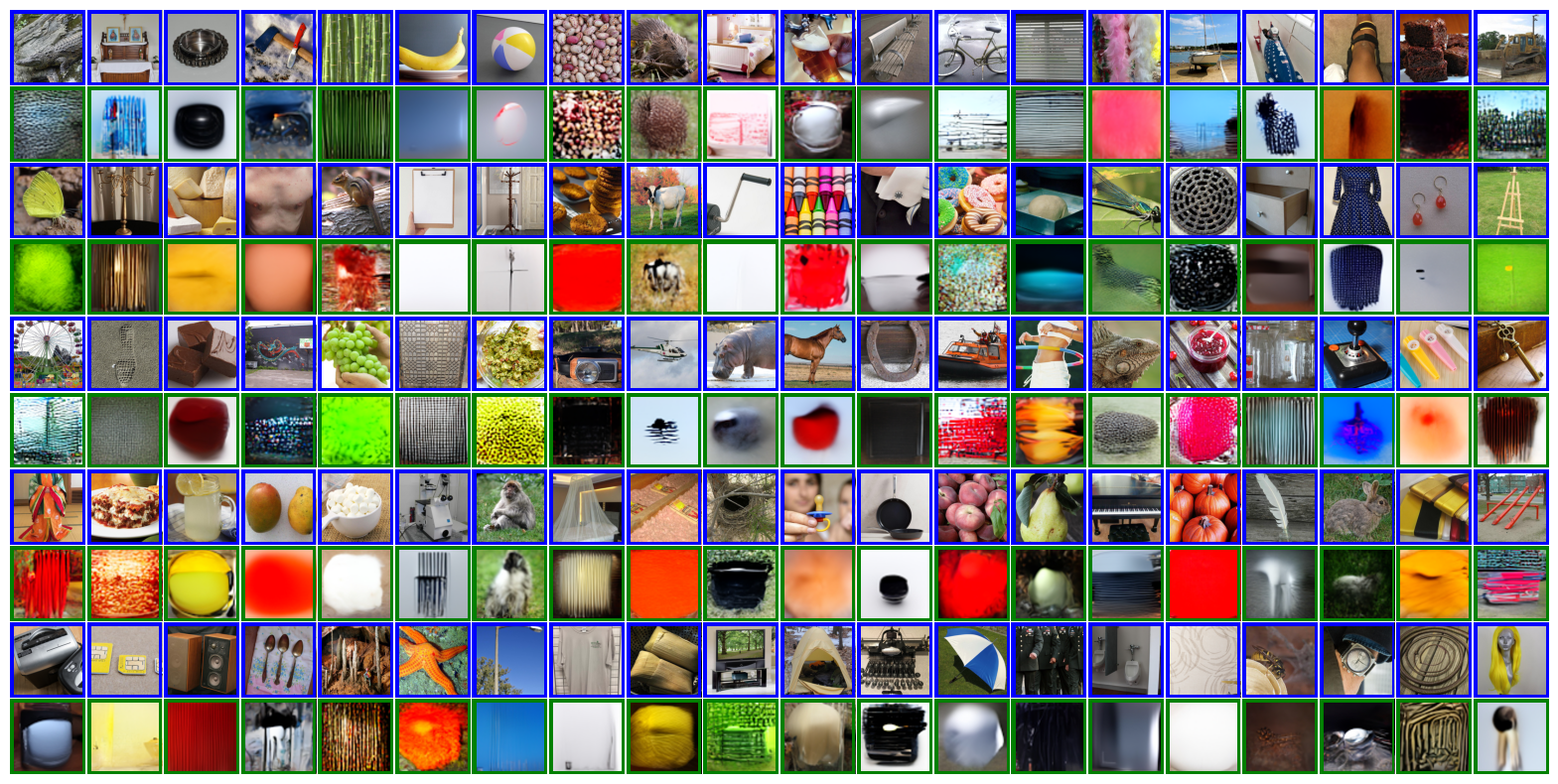}
    \caption{Comparison of ground truth images (blue) with VDVAE reconstructions (green) generated from VDVAE latents decoded from TVSD Monkey F visual cortex spike data.}
    \label{fig:monkey_f_vdvae}
\end{figure}


\begin{figure}
    \centering
    \includegraphics[width=1\linewidth]{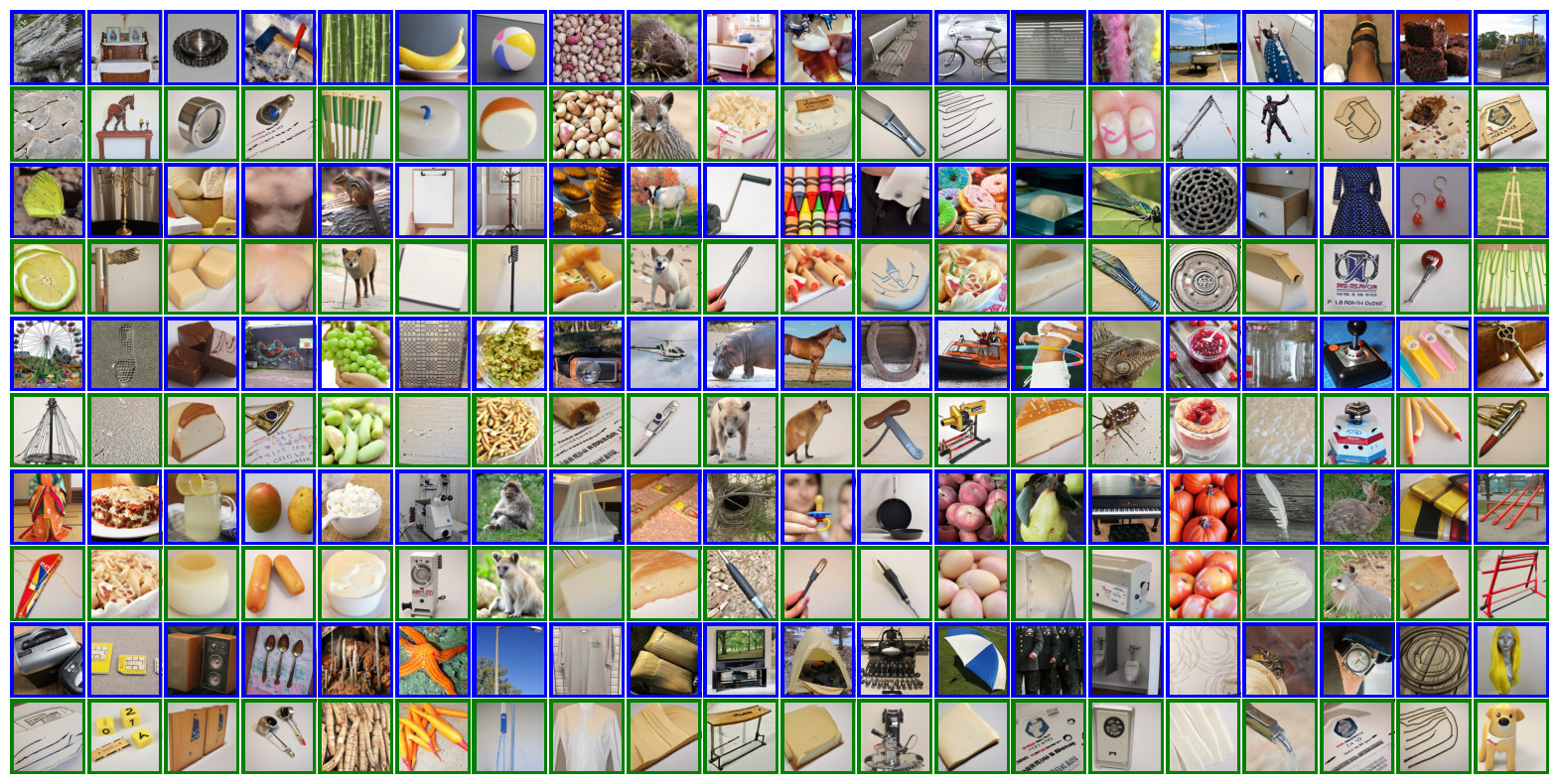}
    \caption{Comparison of ground truth images (blue) with unCLIP reconstructions (green) generated from CLIP embeddings decoded from monkey N visual cortex spike data.}
    \label{fig:monkey_n_unclip}
\end{figure}

\begin{figure}
    \centering
    \includegraphics[width=1\linewidth]{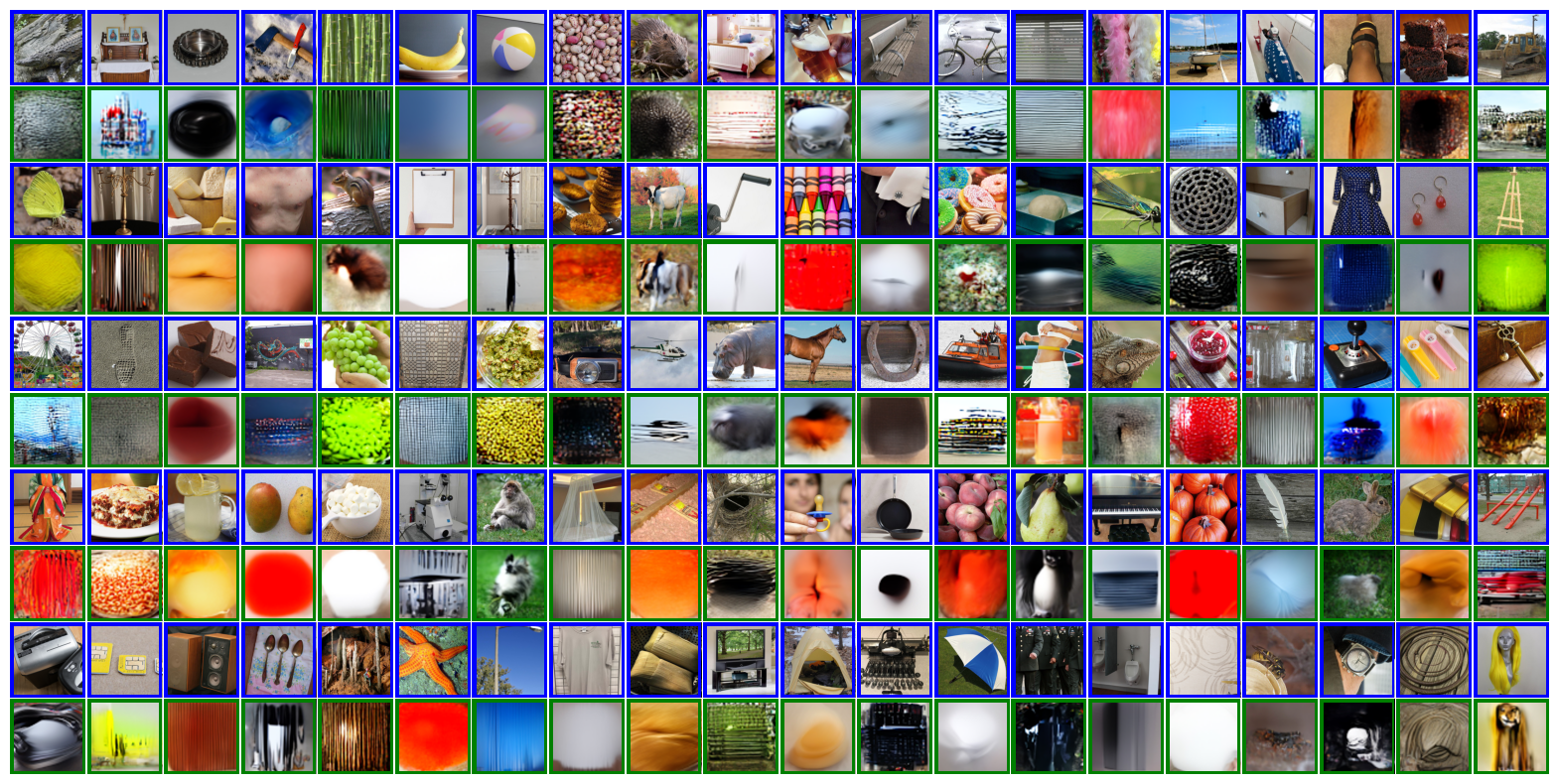}
    \caption{Comparison of ground truth images (blue) with VDVAE reconstructions (green) generated from VDVAE latents decoded from monkey N visual cortex spike data.}
    \label{fig:monkey_n_vdvae}
\end{figure}



\begin{table*}[h] 
\centering
\caption{Quantitative assessments of the reconstruction quality for BM160, Monkey F and Monkey N. 
For detailed explanations of the metrics see section \ref{evaluation_metrics}.}
\label{tab:performance}
\resizebox{\textwidth}{!}{%
\begin{tabular}{lcccccccc}
\toprule
 & \multicolumn{4}{c}{Low-level} & \multicolumn{4}{c}{High-level} \\
\cmidrule(r){2-5} \cmidrule(l){6-9}
Dataset  & PixCorr $\uparrow$ & SSIM $\uparrow$ & AlexNet(2) $\uparrow$ & AlexNet(5) $\uparrow$ & Inception $\uparrow$ & CLIP $\uparrow$ & EffNet $\downarrow$ & SwAV $\downarrow$  \\
\midrule
TITD - BM160 (Perceptogram) & $0.251$ & $0.390$  & $0.891$  & $0.960$ & $0.813$ & $0.812$ & $0.816$ & $0.492$ \\
TVSD - Monkey F (Perceptogram) & $0.214$ & $0.414$  & $0.862$  & $0.921$ & $0.722$ & $0.815$ & $0.859$ & $0.520$ \\
TVSD - Monkey N (Perceptogram) & $0.218$ & $0.417$  & $0.880$  & $0.903$ & $0.734$ & $0.818$ & $0.867$ & $0.538$ \\

\bottomrule
\end{tabular}%
}
\label{tab:performance}
\end{table*}


\subsection{Latent-Filtered Spike Patterns Reveal Distinct Spatiotemporal Dynamics of Feature Encoding in IT Cortex}

\subsubsection{CLIP-Filtered Patterns}
We applied our bidirectional encoding-decoding framework to isolate neural activity components that specifically covary with CLIP semantic representations in macaque IT cortex (BM160, TITD dataset). The resulting spatiotemporal maps revealed category-specific differences. Here we present the results from BM160.

\textbf{Animal stimuli} 
elicited the most robust and spatially focused responses, with activation strongly concentrated at the very top of the most anterior array (top block). These responses emerged around 125-150ms post-stimulus and maintained sustained activation through later time bins. The concentrated anterior localization suggests specialized neural populations for processing animacy-related stimuli in anterior IT.

\textbf{Food stimuli} 
produced more spatially distributed patterns as early as 75-100ms mainly across the middle two electrode arrays. The earlier onset and more scattered distribution compared to animal responses suggests different processing dynamics for this category, potentially reflecting the more heterogeneous visual features within food stimuli.

\subsubsection{VDVAE-Filtered Patterns}
To examine texture encoding independent of semantic content, we applied VDVAE filtering to neural responses, grouping test images by their spatial frequency content computed from Fourier magnitude spectra.

Both \textbf{smooth} (low spatial frequency) and \textbf{textured} (high spatial frequency) stimuli 
showed scattered activation patterns primarily in the middle two electrode arrays, with responses emerging early around 75-100ms post-stimulus. The smooth stimuli showed slightly more activation extending into the bottom (most posterior) array. The similar onset timing but different spatial distributions suggest that IT cortex processes spatial frequency information through partially distinct neural populations.

\begin{figure}
    \centering
    \includegraphics[width=1\linewidth]{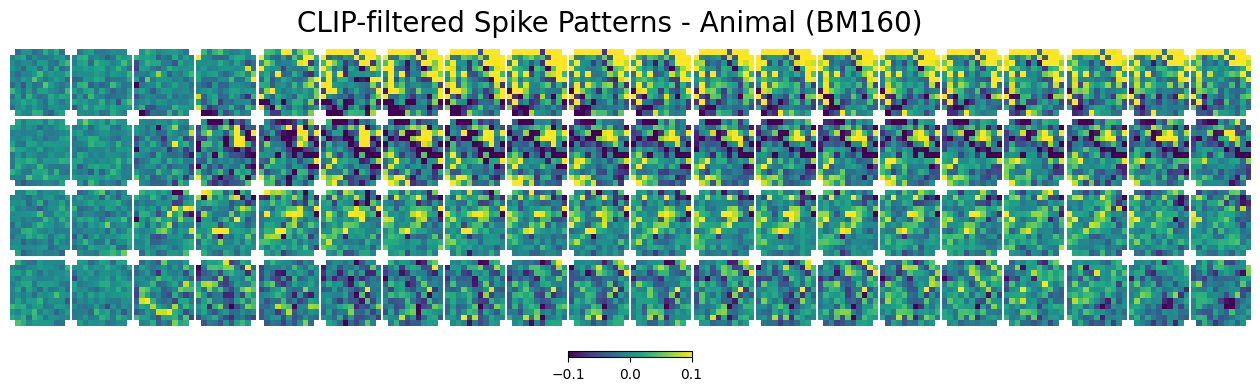}
    \caption{\textbf{CLIP-filtered spike patterns for animal category stimuli.} Spatiotemporal activity patterns showing neural responses that covary with CLIP latent representations, obtained by passing decoded CLIP predictions through encoding weights. Data are averaged across test images clustered as animals using hierarchical clustering of CLIP latents. Each column represents a 25 ms time bin from 0-600 ms post-stimulus onset. Four horizontal blocks correspond to electrode arrays positioned from anterior (top) to posterior (bottom) IT cortex, with electrodes within each block arranged according to their physical array positions.}
    \label{fig:clip_patterns_animal_bm160}
\end{figure}

\begin{figure}
    \centering
    \includegraphics[width=1\linewidth]{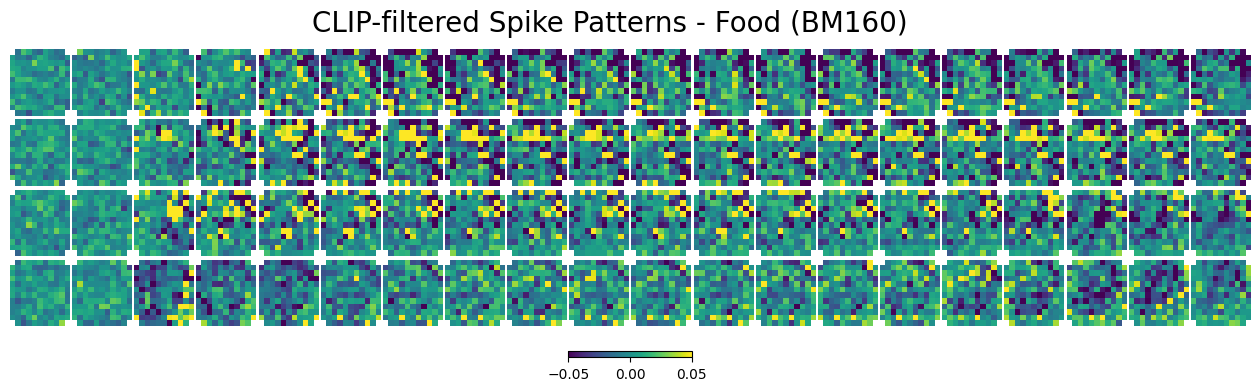}
    \caption{\textbf{CLIP-filtered spike patterns for food category stimuli.} Spatiotemporal activity patterns showing neural responses that covary with CLIP latent representations, obtained by passing decoded CLIP predictions through encoding weights. Data are averaged across test images clustered as animals using hierarchical clustering of CLIP latents. Each column represents a 25 ms time bin from 0-600 ms post-stimulus onset. Four horizontal blocks correspond to electrode arrays positioned from anterior (top) to posterior (bottom) IT cortex, with electrodes within each block arranged according to their physical array positions.}
    \label{fig:clip_patterns_food_bm160}
\end{figure}

\subsection{Encoding Weight Reconstructions Visualizes Presumptive Neural Preference of the Ventral Visual Cortex}

To directly visualize what visual features drive neural responses across IT cortex, we reconstructed preferred stimuli from the CLIP-to-spike encoding weights using unCLIP decoding. This approach treats each weight vector as a CLIP embedding, revealing the "ideal stimulus" that would maximally activate each electrode according to our learned linear model.

\begin{figure}
    \centering
    \includegraphics[width=1\linewidth]{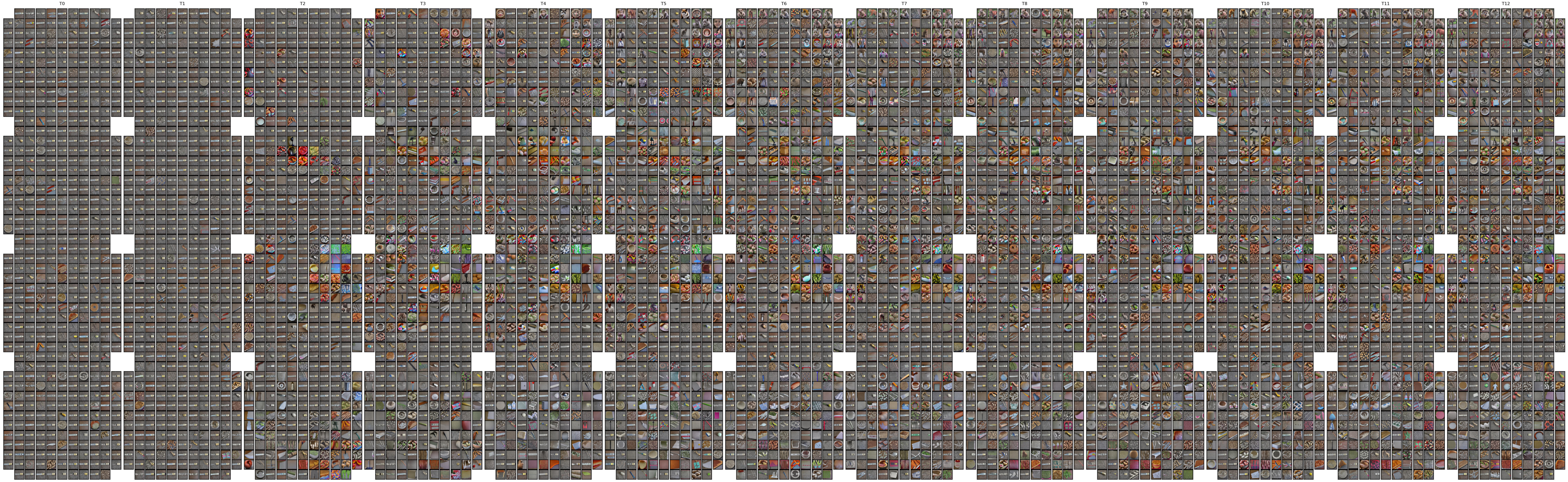}
    \caption{
    \textbf{Spatiotemporal Neural Preference Visualization by performing visual reconstruction on the CLIP-to-spike Neural Encoding Weights of the Monkey BM160's IT Cortex (Please zoom in to see better). } 
    This figure visualizes the preferred stimulus features for hundreds of neural sites across macaque Inferotemporal (IT) cortex over time. Each image is a reconstruction generated by treating a single vector from the neural encoding weights as a CLIP embedding and decoding it with unCLIP. These weights map from the CLIP embedding space to the spike counts of each electrode in each time bin. Thus, each image represents the stimulus that would, according to the linear model, maximally activate a given electrode at a specific moment in time. The figure is organized as a spatiotemporal map: columns represent 25 ms time bins (progressing left-to-right from 0-325 ms), and the four row blocks correspond to four electrode arrays ordered from anterior (top) to posterior (bottom) IT cortex. Within each block, images are arranged according to the electrode's physical location on the array. This visualization shows the evolution from simple feature tuning to more complex object representation over time and across different regions of IT cortex.}
    \label{fig:clip_encoding_weight_bm160}
\end{figure}

The spatiotemporal reconstruction maps for BM160 (Figure \ref{fig:clip_encoding_weight_bm160}) revealed a progression in feature complexity across time and along the anterior-posterior axis of IT. At early timepoints (75-100 ms), reconstructions consisted primarily of simple geometric patterns, color patches, and edge-like features in the posterior (lower) three arrays. These reconstructions are localized into three clusters in the lower three arrays. Around 125 ms, the three clusters start expanding in physical size, and the bottom one goes from complex shapes into simpler shapes. Some face electrodes start to show sparsely at the very anterior (top) of the front (top) array. Around 150ms, the faces electrodes at the top grows into a larger cluster, where it maintains throughout the rest of the time period. 

\begin{figure}
    \centering
    \includegraphics[width=0.8\linewidth]{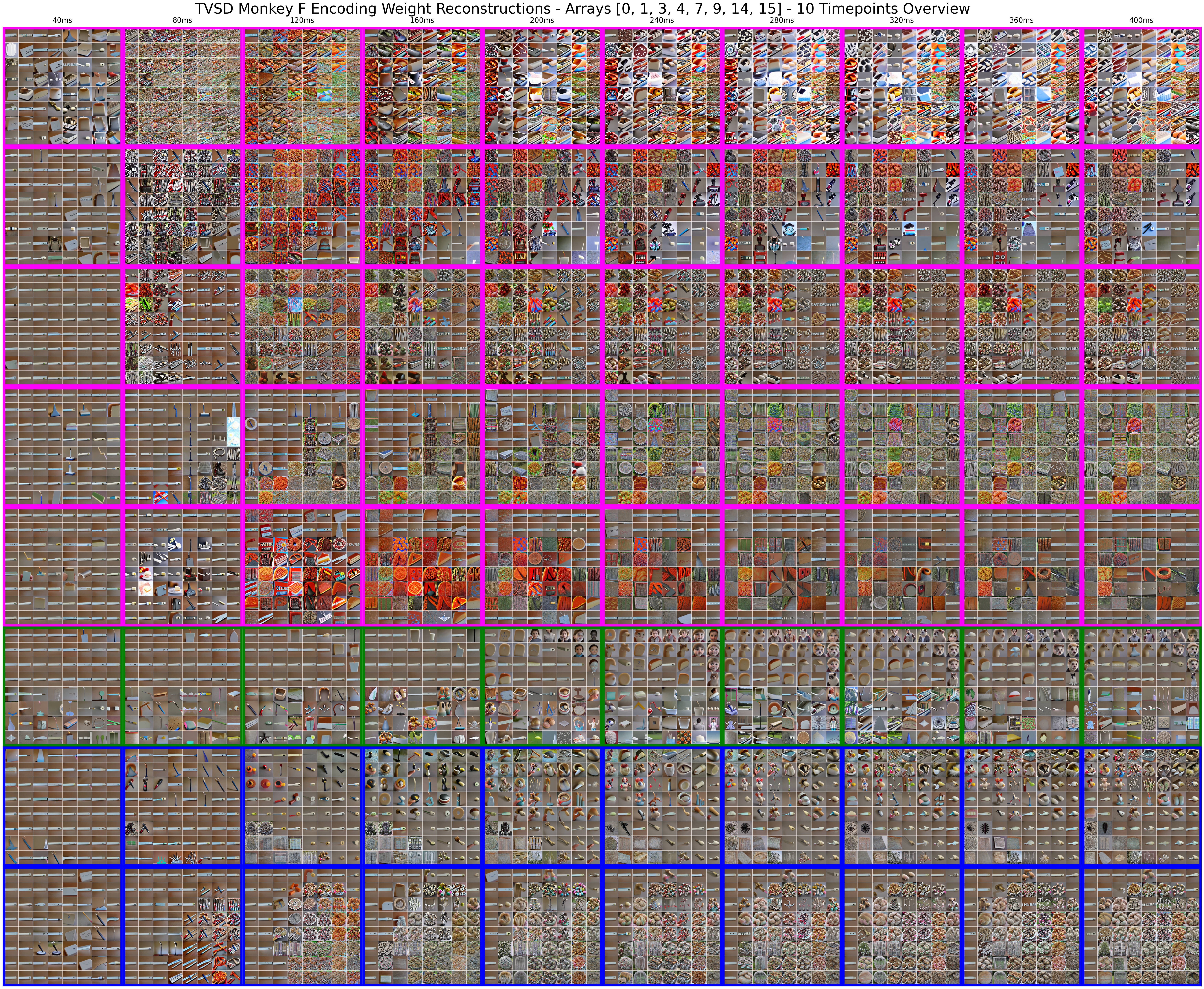}
    \caption{
    \textbf{Spatiotemporal Neural Preference Visualization by performing visual reconstruction on the CLIP-to-spike Neural Encoding Weights of the Monkey F's Ventral Visual Cortex (Please zoom in to see better). } 
    This figure visualizes the preferred stimulus features for hundreds of neural sites across macaque Inferotemporal (IT) cortex over time. Each image is a reconstruction generated by treating a single vector from the neural encoding weights as a CLIP embedding and decoding it with unCLIP. These weights map from the CLIP embedding space to the spike counts of each electrode in each time bin. Thus, each image represents the stimulus that would, according to the linear model, maximally activate a given electrode at a specific moment in time. The figure is organized as a spatiotemporal map: columns represent 40 ms time bins (progressing left-to-right from 0-400 ms), and the 8 row blocks correspond to 8 selected electrode arrays with border color indicating the location 
    (Magenta--V1, Green--IT, Blue--V4)
    . Within each block, images are arranged according to the electrode's physical location on the array. This visualization shows the evolution from simple feature tuning to more complex object representation over time and across different regions of visual cortex. The bottom half of the selected IT array appears to be noisy.}
    \label{fig:clip_encoding_weight_monkey_f}
\end{figure}

The spatiotemporal reconstruction maps for Monkey F (Figure \ref{fig:clip_encoding_weight_monkey_f}) revealed a progression in feature complexity across time and along the array locations. At early timepoints (40-80 ms), the "optimal stimuli" consisted primarily of shapes and textures in V1 arrays. Color features reaches maximum by 120ms in both V1 and V4 and start to decay from there, and the V4 arrays return to texture-related features by 160ms. Some IT electrodes start to show face-like features afterwards (160-200 ms) and continues increasing from there. 

\subsubsection{Weight Similarity Analysis Reveals Three Clusters in Monkey BM160}
\begin{figure}
    \centering
    \includegraphics[width=1\linewidth]{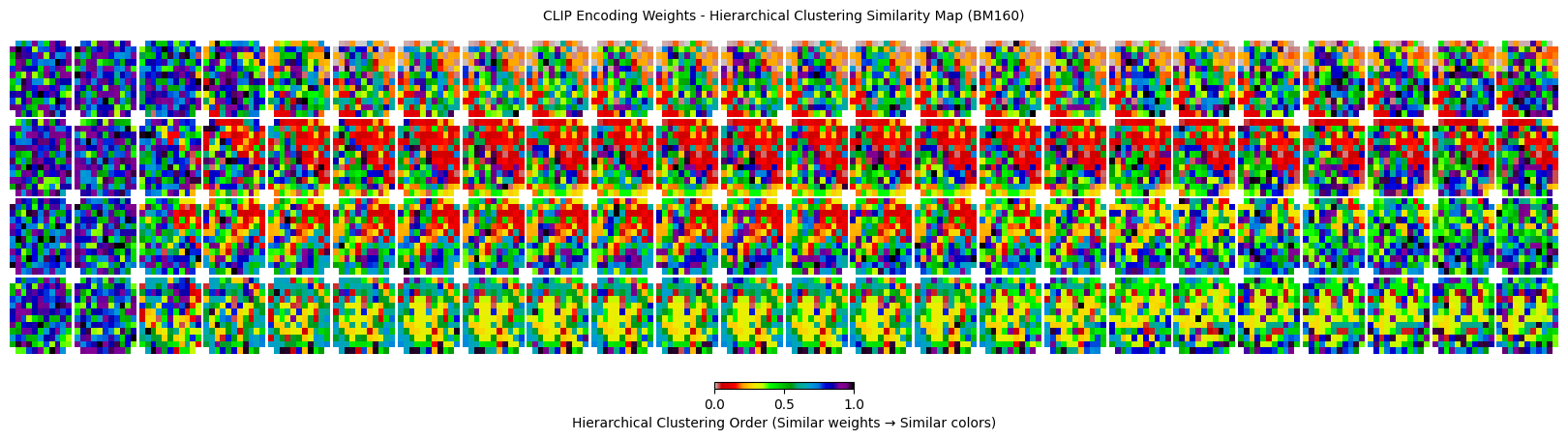}
    \caption{\textbf{Weight Similarity Map of CLIP Preference in Macaque BM160's IT Cortex. } This figure visualizes the functional similarity of neural tuning across hundreds of recording sites and multiple time points. Each vector from the neural encoding weights was treated as a CLIP embedding. They were aggregated and subjected to hierarchical clustering to group them by similarity. This clustering produces a one-dimensional ordering of the weights, which was then mapped to the continuous color scale shown in the color bar. Each pixel in the main figure is colored according to its position in this similarity-based ordering. The figure is organized as a spatiotemporal map: columns represent 25 ms time bins (progressing left-to-right from 0-600 ms), and the four row blocks correspond to four electrode arrays ordered from anterior (top) to posterior (bottom) IT cortex. Within each block, images are arranged according to the electrode's physical location on the array. The emergence of large, contiguous patches of uniform color (e.g., the gray cluster in the top array, the red clusters in the second array, the yellow/green cluster in the bottom array) signifies the formation of functional ensembles of neurons with homogeneous tuning properties.}
    \label{fig:clip_weight_similarity}
\end{figure}

The hierarchical clustering of encoding weights, which is visualized as a similarity map (Figure \ref{fig:clip_weight_similarity}), showed three contiguous patches of uniform color -- a gray cluster in anterior regions, red clusters in the middle two arrays, and a yellow cluster in posterior arrays -- indicating functional ensembles of neurons with homogeneous tuning properties.

\begin{figure}[!htbp]
    \centering
    \begin{subfigure}{0.3\linewidth}
        \centering
        \includegraphics[width=\linewidth]{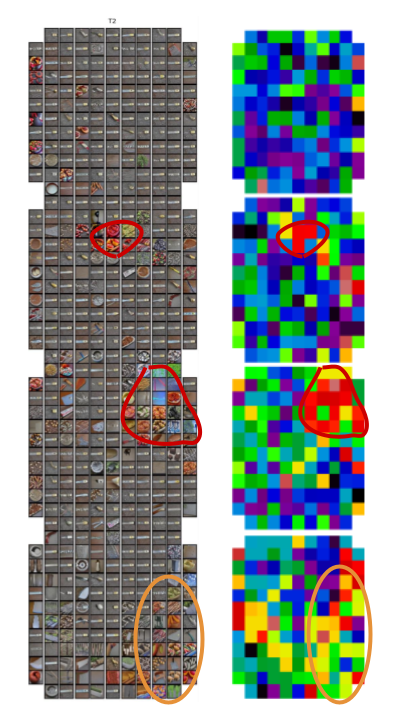}
        \caption{75ms. Small red clusters emerges in the middle two arrays corresponding to mid-level visual features such as color and texture in the weight reconstructions on the left in red circles.}
        \label{fig:weight75}
    \end{subfigure}
    \begin{subfigure}{0.3\linewidth}
        \centering
        \includegraphics[width=\linewidth]{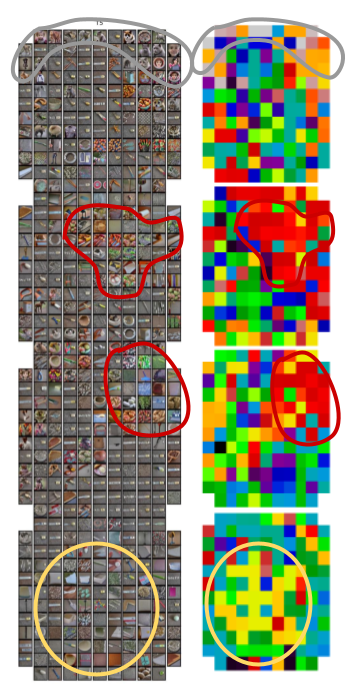}
        \caption{150ms. Gray circles highlight face-selective regions, corresponding to tightly clustered gray patches in the similarity map.}
        \label{fig:weight150}
    \end{subfigure}
    \caption{Weight reconstruction and similarity maps side by side at specific timepoints. Gray circles highlight face-selective regions showing early emergence of face-like features and coherent functional clustering. Red and orange circles mark areas tuned to mid-level visual features such as color patches and texture patterns, while the yellow circle indicates a posterior region selective for simpler geometric shapes.}
    \label{fig:weight_visualization}
\end{figure}

The similarity map corresponds strongly with the reconstruction map. Figure \ref{fig:weight_visualization} illustrates this with maps from two time bins -- one from 50-75ms and one from 125-150ms. The red circle areas appears in the early time bin as a small area. The orange circle shows it's mainly yellow electrodes with some other colors mixed with it, and from the reconstruction map it shows more mid-level features such as texture than the yellow circle later on in the same area. It also shows that the gray area corresponds to the face area in the reconstruction map and only appears later on, indicating temporal processing hierarchy.



\section{Discussion}

In this study, we present a new framework for interpreting monkey visual processing by leveraging both encoding and decoding approaches; and both simplicity of linear model and the power of vision foundation models on large-scale neural recordings from the 
TITD and 
TVSD datasets. Our results provide new insights into the spatiotemporal dynamics and functional organization of the primate ventral visual stream, and demonstrate the feasibility of reconstructing perceptual content, particularly semantically-related, from distributed neural activity.

\subsection{Decoding Visual Percepts from Population Activity}

Our decoding pipeline, which maps multi-electrode spiking activity to latent visual representations and reconstructs images using generative models, demonstrates that it is possible to recover both low-level (e.g., color, texture) and high-level (e.g., semantic category) features of visual stimuli from population activity in the monkey ventral stream. Reconstructions from both the 
TITD and 
TVSD datasets were consistent with the original stimuli, as quantified by metrics such as pixel correlation, SSIM, and CLIP-based feature similarity. Visually, reconstructions from CLIP latents preserved semantic categories, while VDVAE-based reconstructions emphasized lower-level features such as color and texture. Our results show that a linear projection from spiking activity to visual latent spaces is a simple yet effective way of extracting visual representations from distributed neural codes.

\subsection{Encoding Models Reveals Spatiotemporal Dynamics and Functional Organization}

Our use of encoding models to map visual features back to neural activity enables the visualization of "preferred stimuli" for each electrode, providing interpretable representations of neural tuning. This is an advantage provided by the simplicity of linear projection combined with the power of visual foundation model. By analyzing the spatiotemporal evolution of neural responses, we reveal distinct patterns of feature encoding across the IT cortex. For example, animal stimuli elicited robust, spatially focused responses in anterior IT, emerging around 125–150 ms post-stimulus, while food stimuli produced more distributed patterns that appeared earlier in middle IT regions. Similarly, textured and smooth stimuli evoked distinct activation patterns, highlighting the cortex’s sensitivity to spatial frequency. These results are consistent with the hierarchical organization of the ventral stream, where posterior regions encode simple features and anterior regions encode complex objects and categories \cite{Tanaka1996, Gross1992}.

The use of both the TITD and TVSD datasets underscores the generalizability of our approach. The TITD dataset, with its dense coverage of IT cortex, allows for detailed mapping of functional organization, while the TVSD dataset provides broader coverage across V1, V4, and IT, enabling the study of hierarchical processing across the ventral stream. 
The consistency of our results across these datasets highlights the robustness of our methods and the value of large-scale, naturalistic neural recordings for advancing our understanding of visual perception.



\section{Funding Information}

This research was supported in part by the National Institute of Information and Communications Technology (NICT) grant NICT 22301(RH), and the Japan Science and Technology Agency, Moonshot Research \& Development Program grant JPMJMS2012 (RH and HN) and MEXT/JSPS KAKENHI Grantin-Aid for Transformative Research Areas (A), Grant Number 24H02185 (RH) and Grant-in-Aid for Scientific Research (B), Grant Number 24K03241 (RH) and Grant-in-Aid for Early-Career Scientists 24K20883 (HN).The animal used in the TITD dataset was provided by the National BioResource Project (Japan).  This work was also supported in part by NSF IIS 1817226, CRCNS 2208362, and a seed grant from UC San Diego Social Sciences. The authors are also grateful for the hardware support provided by NVIDIA, Adobe, and Sony.

\bibliographystyle{apalike}    
\bibliography{references}  

\newpage

\appendix
\counterwithin{figure}{section}





\section{Extended Results}

\begin{figure}[h]
    \centering
    \includegraphics[width=1\linewidth]{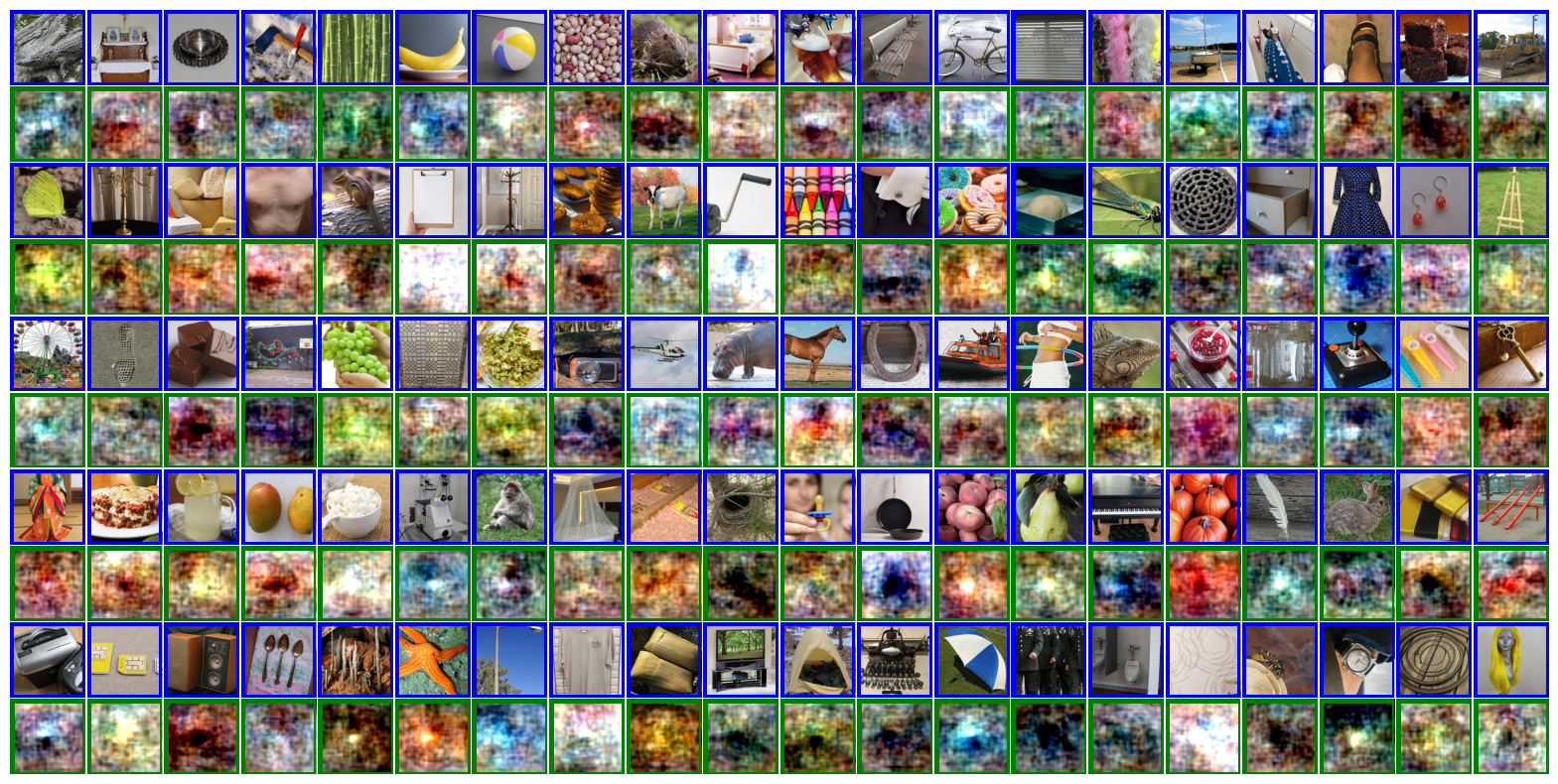}
    \caption{Comparison of ground truth images (blue) with PCA reconstructions (green) generated from PCA components decoded from TITD Monkey BM160 IT cortex spike data.}
    \label{fig:pca1k_25msbin}
\end{figure}

\begin{figure}
    \centering
    \includegraphics[width=1\linewidth]{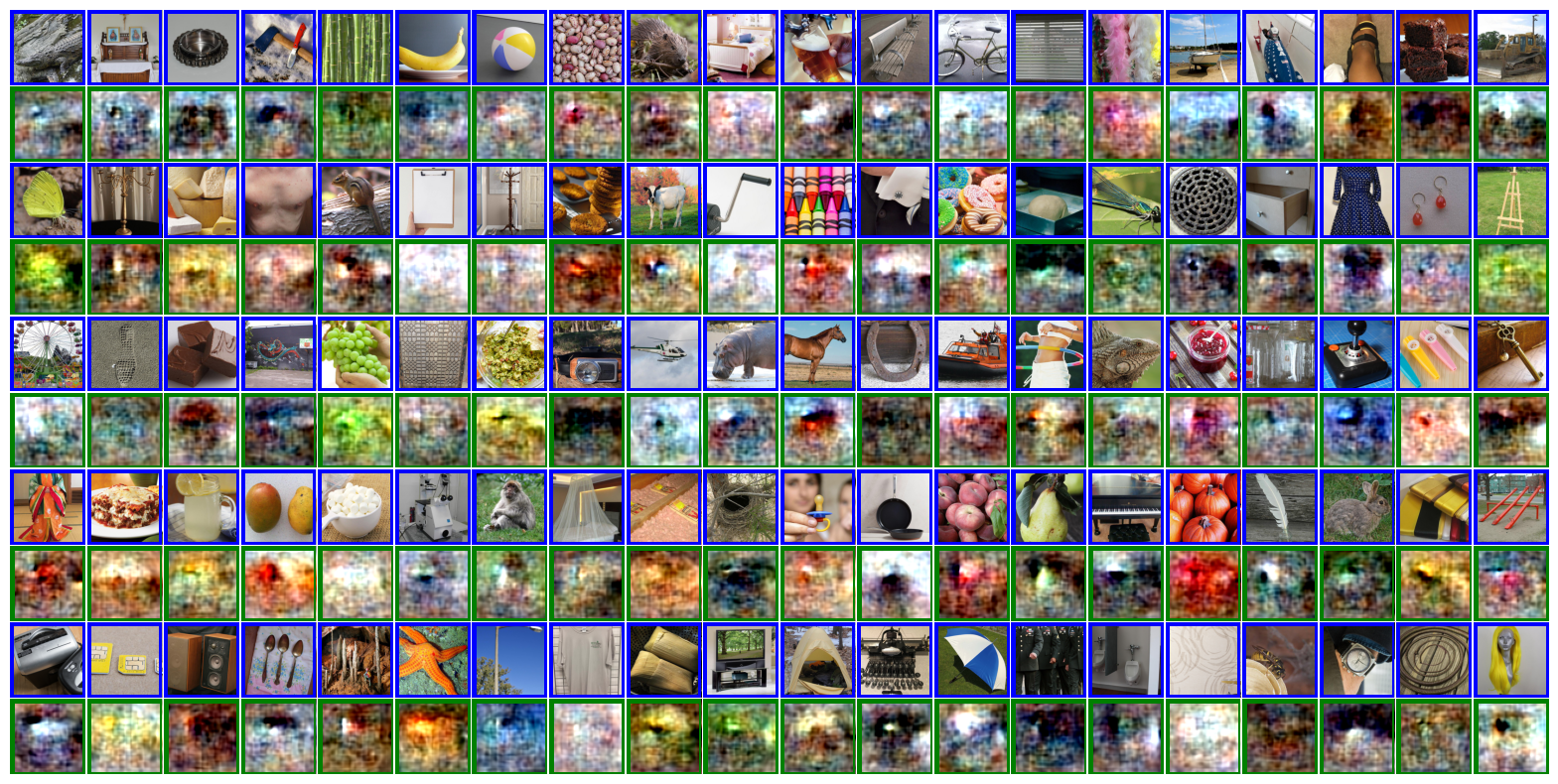}
    \caption{Comparison of ground truth images (blue) with PCA reconstructions (green) generated from PCA components decoded from TVSD Monkey F visual cortex spike data.}
    \label{fig:monkey_f_pca1k}
\end{figure}

\begin{figure}
    \centering
    \includegraphics[width=1\linewidth]{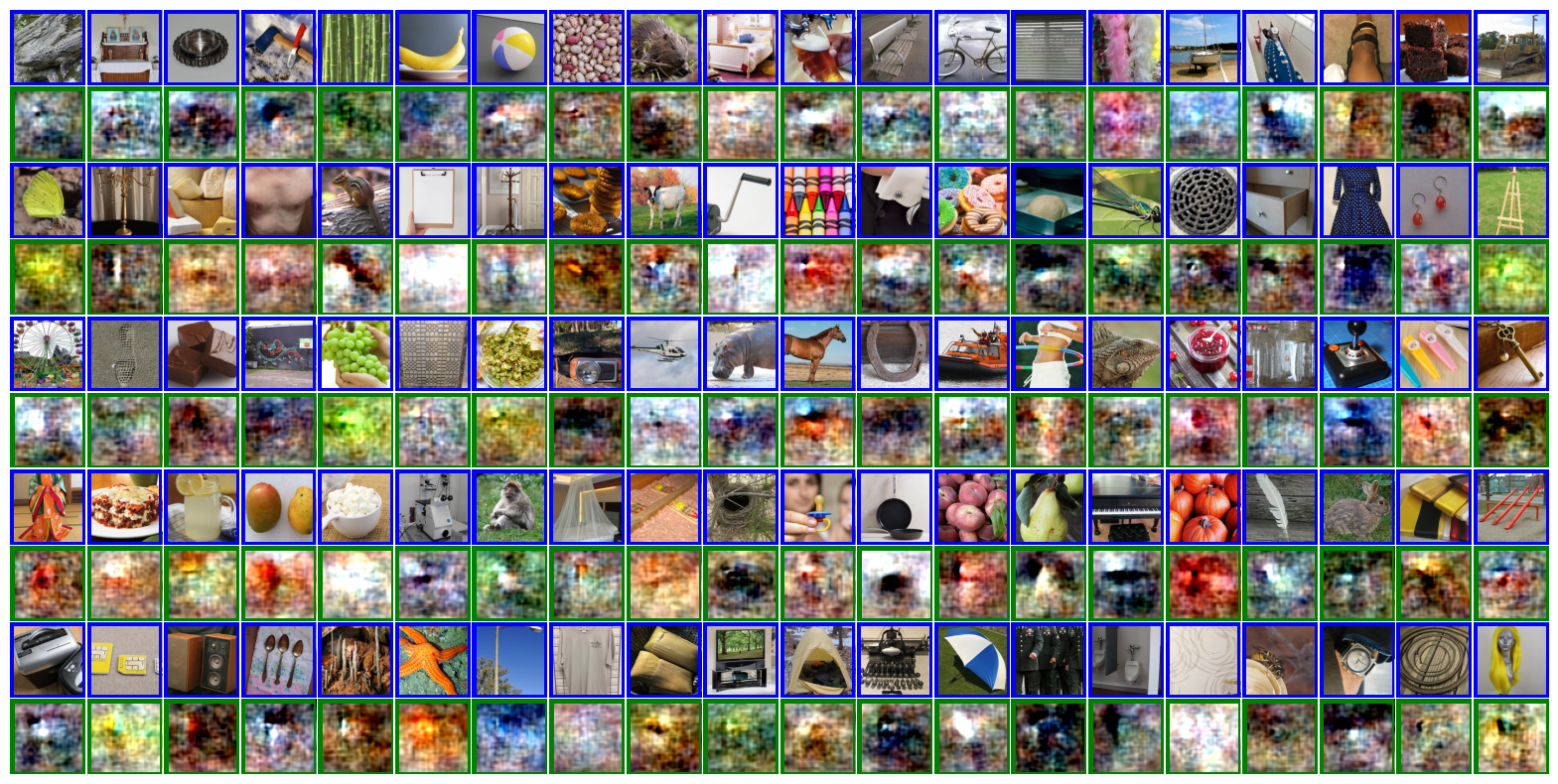}
    \caption{Comparison of ground truth images (blue) with PCA reconstructions (green) generated from PCA components decoded from TVSD Monkey N visual cortex spike data.}
    \label{fig:monkey_n_pca1k}
\end{figure}

\begin{figure}[!htbp]
    \centering

    \begin{subfigure}{0.45\linewidth}
        \centering
        \includegraphics[width=\linewidth]{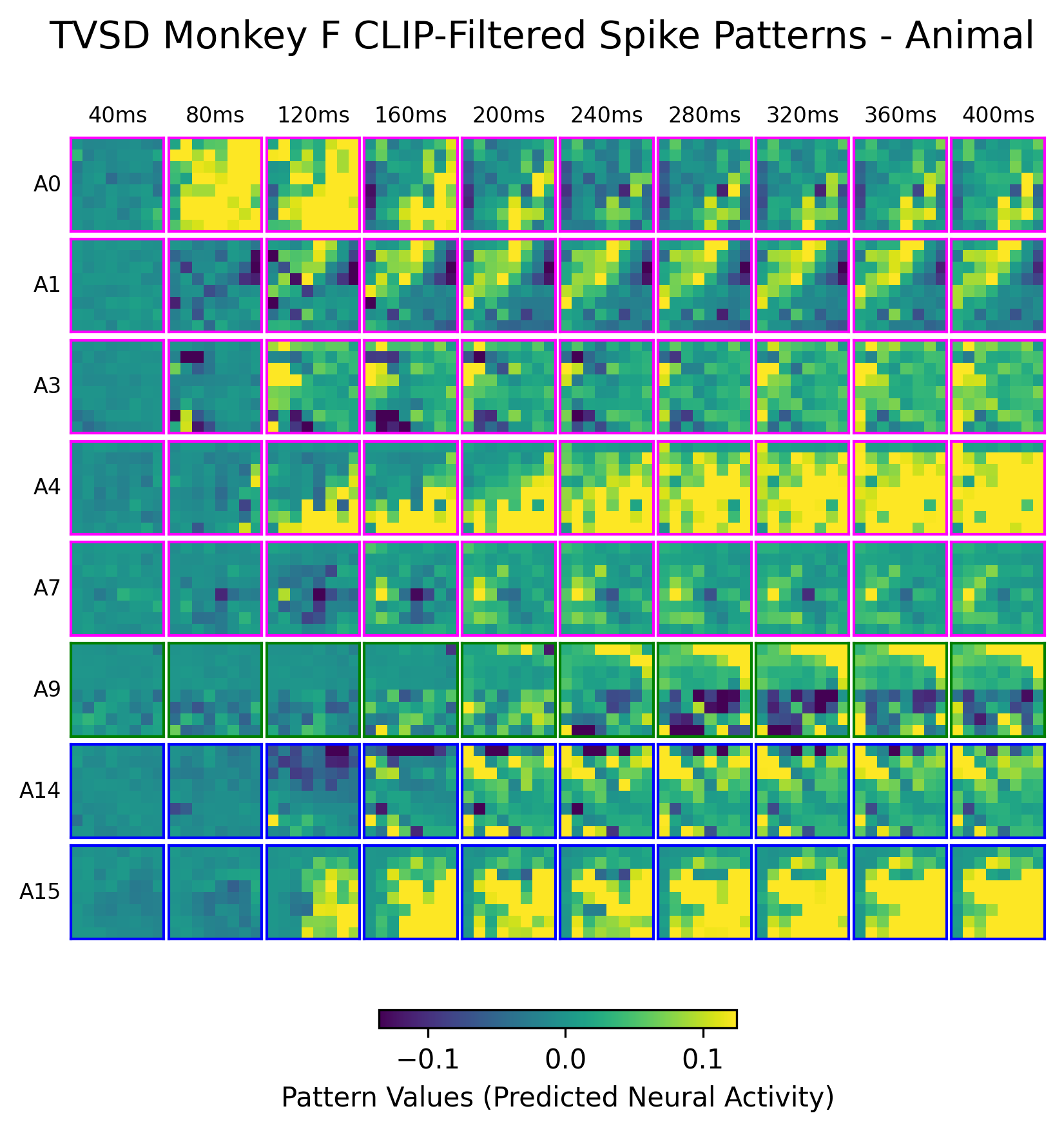}
        \caption{\textbf{CLIP-filtered spike patterns for animal category stimuli.} }
        \label{fig:clip_patterns_animal_f}
    \end{subfigure}
    \hfill
    \begin{subfigure}{0.45\linewidth}
        \centering
        \includegraphics[width=\linewidth]{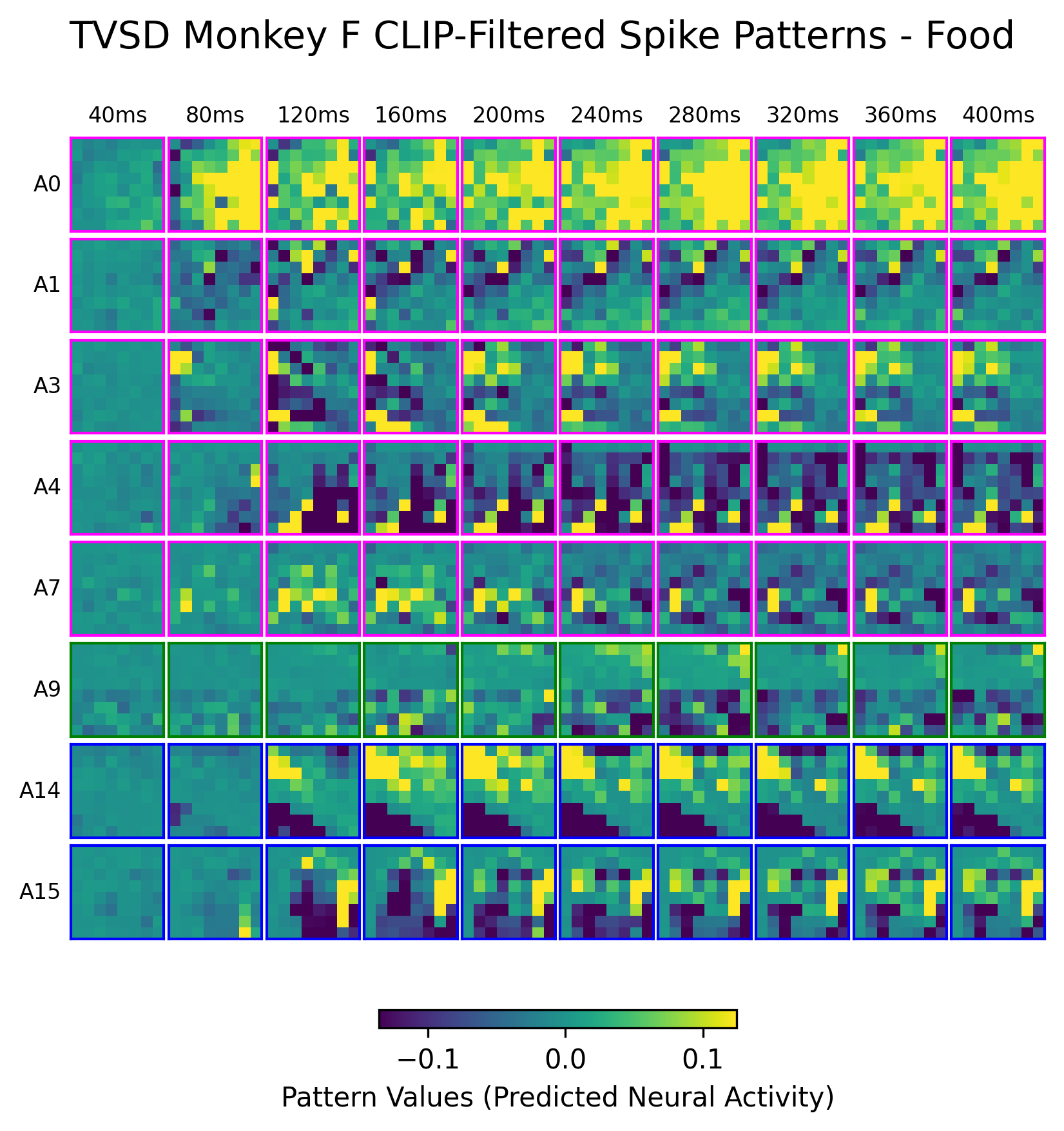}
        \caption{\textbf{CLIP-filtered spike patterns for food category stimuli.} }
        \label{fig:clip_patterns_food_f}
    \end{subfigure}
    \caption{Spatiotemporal activity patterns from TVSD Monkey F visual cortex showing neural responses that covary with CLIP latent representations, obtained by passing decoded CLIP predictions through encoding weights. Data are averaged across test images clustered as animals using hierarchical clustering of CLIP latents. Each column represents a 40 ms time bin from 0-400 ms post-stimulus onset. Eight vertical blocks correspond to electrode arrays in different locations, with the border color indicating the brain region 
    (Magenta--V1, Green--IT, Blue--V4)
    , with electrodes within each block arranged according to their physical array positions.}
    \label{fig:clip_patterns_f}
\end{figure}

\begin{figure}
    \centering
    \includegraphics[width=1\linewidth]{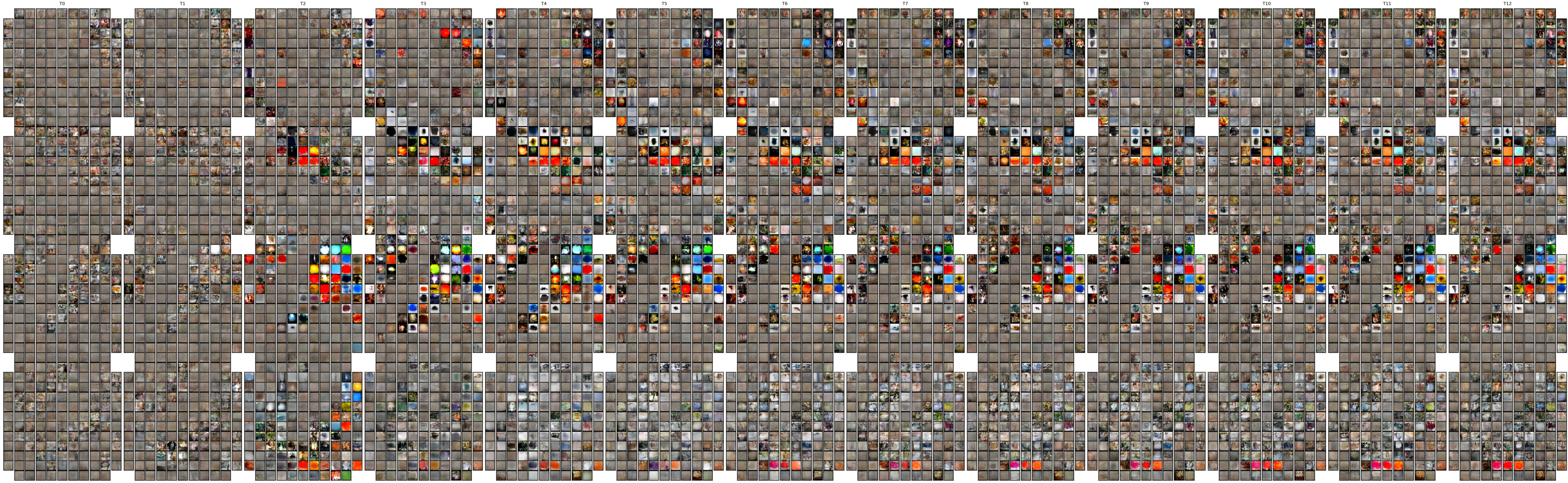}
    \caption{
    \textbf{Spatiotemporal Neural Preference Visualization by performing visual reconstruction on the VDVAE-to-spike Neural Encoding Weights of the Monkey BM160's IT Cortex (Please zoom in to see better). } 
    This spatiotemporal map shows the preferred visual stimulus based on the VDVAE latent space for each electrode (inferred from model weights) over time. Columns represent 25 ms time bins, and row blocks correspond to different recording arrays in IT cortex, illustrating the dynamic evolution of neural tuning.}
    \label{fig:placeholder}
\end{figure}

\begin{figure}
    \centering
    \includegraphics[width=1\linewidth]{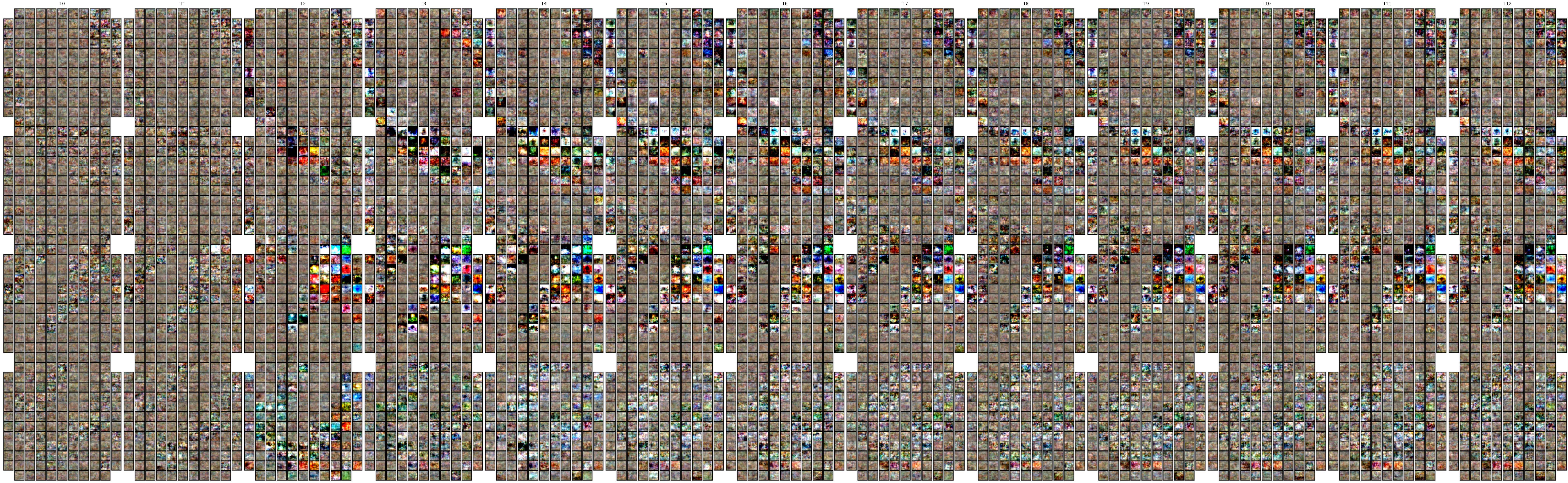}
    \caption{
    \textbf{Spatiotemporal Neural Preference Visualization by performing visual reconstruction on the PCA-to-spike Neural Encoding Weights of the Monkey BM160's IT Cortex (Please zoom in to see better). } 
    This spatiotemporal map shows the preferred visual stimulus based on the PCA latent space for each electrode (inferred from model weights) over time. Columns represent 25 ms time bins, and row blocks correspond to different recording arrays in IT cortex, illustrating the dynamic evolution of neural tuning.}
    \label{fig:placeholder}
\end{figure}

\begin{figure}
    \centering
    \includegraphics[width=0.8\linewidth]{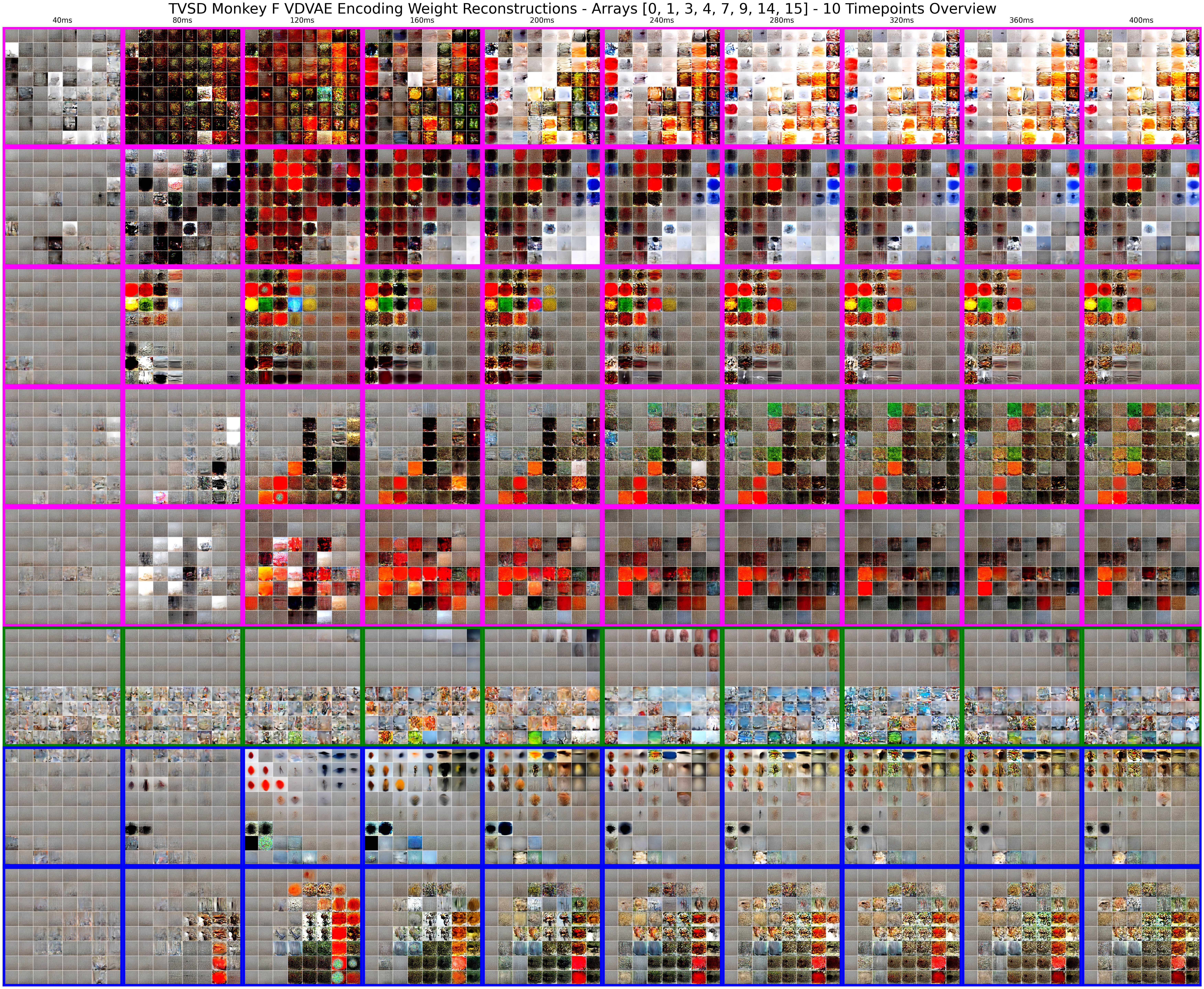}
    \caption{
    \textbf{Spatiotemporal Neural Preference Visualization by performing visual reconstruction on the VDVAE-to-spike Neural Encoding Weights of the Monkey F's Ventral Visual Cortex (Please zoom in to see better). } 
    This spatiotemporal map shows the preferred visual stimulus based on the VDVAE latent space for each electrode (inferred from model weights) over time. Columns represent 40 ms time bins, and row blocks correspond to different recording arrays, illustrating the dynamic evolution of neural tuning.}
    \label{fig:vdvae_encoding_weight}
\end{figure}

\begin{figure}[!htbp]
    \centering
    \includegraphics[width=0.8\linewidth]{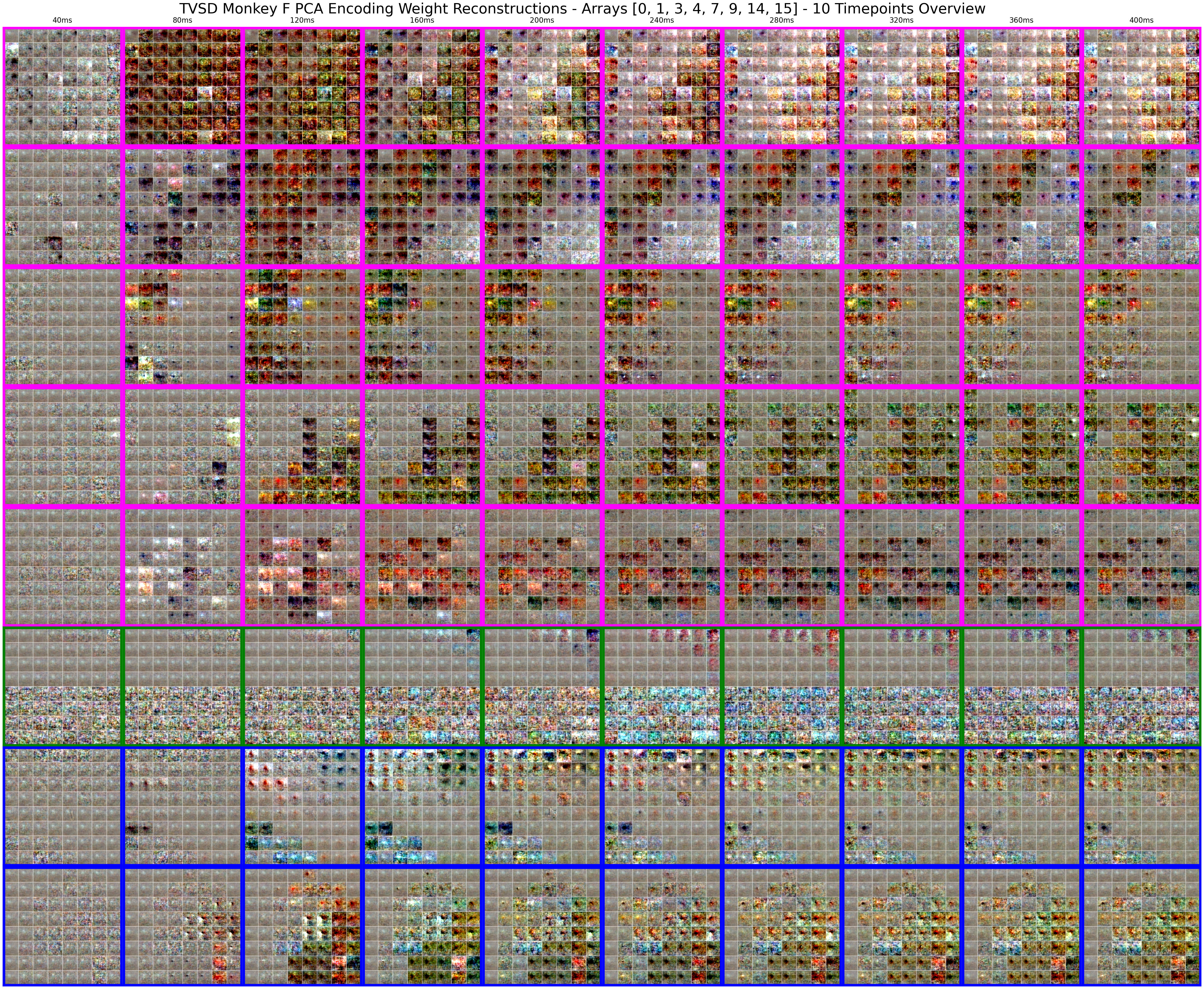}
    \caption{
    \textbf{Spatiotemporal Neural Preference Visualization by performing visual reconstruction on the PCA-to-spike Neural Encoding Weights of the Monkey F's Ventral Visual Cortex (Please zoom in to see better). } 
    This spatiotemporal map shows the preferred visual stimulus based on the PCA latent space for each electrode (inferred from model weights) over time. Columns represent 40 ms time bins, and row blocks correspond to different recording arrays, illustrating the dynamic evolution of neural tuning.}
    \label{fig:pca_encoding_weight}
\end{figure}

\end{document}